\documentclass[fleqn,10pt]{wlpeerj}

\usepackage{fullpage}

\usepackage[utf8]{inputenc}
\usepackage{multirow}
\usepackage{color}

\usepackage{amssymb,amsmath}
\usepackage{graphicx}
\usepackage{booktabs}
\usepackage{subfig}
\usepackage{listings, xcolor}

\usepackage{xcite}

\usepackage{xr-hyper}
\usepackage{hyperref}

\usepackage{cleveref}

\makeatletter
\newcommand*{\addFileDependency}[1]{% argument=file name and extension
  \typeout{(#1)}
  \@addtofilelist{#1}
  \IfFileExists{#1}{}{\typeout{No file #1.}}
}
\makeatother
\newcommand*{\myexternaldocument}[1]{%
    \externaldocument{#1}%
    \addFileDependency{#1.tex}%
    \addFileDependency{#1.aux}%
}
\myexternaldocument{supplementary_material}

\usepackage{longtable}

\DeclareMathOperator*{\argmax}{arg\,max}

\newcommand{\nomopy}{\emph{NoMoPy }}

\newcommand{\arxiv}[1]{}
\newcommand{\peerj}[1]{#1}

\let\oldequation\equation
\let\oldendequation\endequation
\renewenvironment{equation}
  {\linenomathNonumbers\oldequation}
  {\oldendequation\endlinenomath}
\let\oldalign\align
\let\oldendalign\endalign
\renewenvironment{align}
  {\linenomathNonumbers\oldalign}
  {\oldendalign\endlinenomath}

\definecolor{verbgray}{gray}{0.9}
\newenvironment{tcolorbox}{}{}

\title{\textbf{NoMoPy: Noise Modeling in Python}}

\arxiv{
\author{Dylan Albrecht}
\email[Corresponding author ]{dalbrec@sandia.gov}
\address{Sandia National Laboratories, Albuquerque, NM 87185, USA}
\author{N. Tobias Jacobson}
\email[Corresponding author ]{ntjacob@sandia.gov}
\address{Center for Computing Research, Sandia National Laboratories, Albuquerque, NM 87185, USA}
}
\peerj{
\author[1]{Dylan Albrecht}
\author[2]{N. Tobias Jacobson}
\affil[1]{Sandia National Laboratories, Albuquerque, NM 87185, USA}
\affil[2]{Center for Computing Research, Sandia National Laboratories, Albuquerque, NM 87185, USA}
\corrauthor[1]{Dylan Albrecht}{dalbrec@sandia.gov}
}

\begin{abstract}
\nomopy is a code for fitting, analyzing, and generating noise modeled as a hidden Markov model (HMM) or, more generally, factorial hidden Markov model (FHMM). This code, written in Python, implements approximate and exact expectation maximization (EM) algorithms for performing the parameter estimation process, model selection procedures via cross-validation, and parameter confidence region estimation. Here, we describe in detail the functionality implemented in \nomopy and provide examples of its use and performance on example problems.
\end{abstract}

\begin{document}

\arxiv{
\maketitle
\tableofcontents
}
\peerj{
\flushbottom
\maketitle
\thispagestyle{empty}
}

\section{Introduction}
\subsection{Motivation}
The development of \nomopy was prompted by a need to analyze non-Gaussian stochastic time series that may have been generated by a hidden Markov model (HMM) or, more generally, a factorial hidden Markov model (FHMM). In particular, we are interested in systems for which the observed signal is a continuously-distributed function of discrete underlying hidden states that evolve in time according to a stationary Markov process.  Random signals that may be modeled in this form arise frequently, in cases as diverse as the discrete charge fluctuations observed in solid-state electronic devices \cite{Zimmerman1997,Miki2012, Puglisi2014}, magnetic noise in semiconductors \cite{Hensen2020}, sequence analysis in biophysics and bioinformatics \cite{Durbin1998,Husmeier2005}, and energy disaggregation \cite{Kolter2012,Wang2018}. Our goal with \nomopy is to provide an easy to use platform for others to perform this type of analysis by making use of \nomopy\!'s implementations of model fitting, model selection, and parametric uncertainty quantification methods.

% protein and nucleic acid dynamics in biophysics

\subsection{Implemented features}
\nomopy includes implementations of several expectation-maximization (EM) algorithms for FHMMs, including the exact, mean-field, and Gibbs-sampling EM algorithms of Ghahramani and Jordan \cite{Ghahramani1997}. For inference of the hidden state trajectory that is most consistent with the observed time series for a given set of model parameters, we have also implemented the Viterbi algorithm \cite{Viterbi1967} for FHMMs~\cite{Nefian2002,Rabiner1989}.

In addition to our implementations of these published fitting algorithms, we have
incorporated new machinery in \nomopy for performing model selection and confidence
region estimation, including a novel derivation and implementation of analytic
Hessian-based confidence regions for FHMMs. For model selection, we provide a
straightforward
process flow for performing cross-validation on models of interest to test their
performance on data that have not been used for parameter optimization. By considering
models of increasing complexity, this provides a means of identifying a minimum number
of model degrees of freedom that adequately describe the data. Given a model fit, we also
facilitate bootstrapping-based methods for estimating confidence regions
for model parameters.

\section{Related work}

%Application of FHMMs to the analysis of electrical noise in solid-state devices has been reported in \cite{Miki2012, Puglisi2014}.

The only publicly available implementation and the most closely related work we have
found is the \lstinline|factorial_HMM| Python code
of Schweiger et.~al.~\cite{Schweiger2019}.  They present an implementation of the exact
EM algorithm from Ref.~\cite{Ghahramani1997}, extending to include the Viterbi algorithm and
other standard HMM algorithms, as well as addressing the cases of discrete observables,
differing number of states per chain, and time varying transition matrices per chain.
They do not address model selection, confidence regions, nor do they provide implementations
of approximate expectation algorithms (mean field, SVA, and Gibbs), as we do here.

Another related code is present in NILMTK~\cite{Batra2014}.  They have methods for
energy disaggregation using FHMMs, however they do not implement the EM and other
such algorithms.  The NILMTK algorithms are typically used in a supervised learning
setting where the appliance (hidden chains) switching rates are known, or previously
estimated.
\section{Overview of capabilities}
\subsection{Factorial Hidden Markov Models}
\begin{figure}[t]
  \centering
  \includegraphics[scale=0.8]{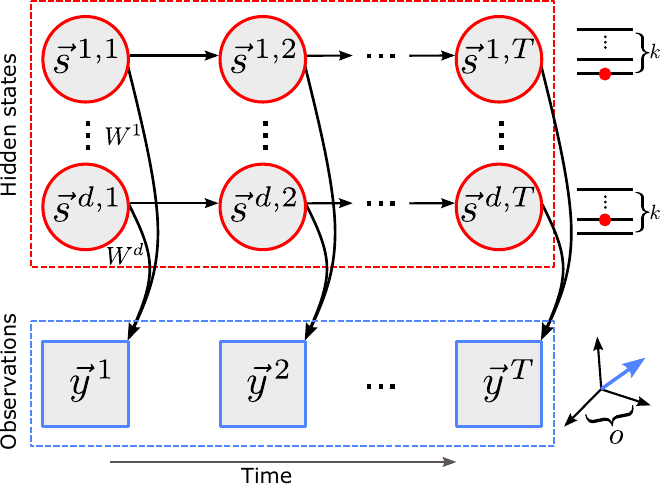}
  \caption{{\bf Graphical representation of the Factorial Hidden Markov Model}.
    The observations, the squares in the bottom row of the graph, are each represented by an $o$-length vector $\vec{y}^{\,t}$ for a total number of $T$ discrete time steps.  These observables depend on the values of $d$ hidden states $\vec{s}^{\,d, t}$, where each hidden state, contained in a circle, can have $k$ values and is represented as a $k$-length vector with 1 in a single entry and 0's in all the others.  Each hidden state depends only on the hidden state at the previous time step, within the same chain, as indicated by the arrows.
}
  \label{fig:fhmm_diagram}
\end{figure}
FHMMs are used to model vector
time series data, with (hidden) variable dependence shown in Fig.~\ref{fig:fhmm_diagram}.
The observable denoted $\vec{y}^{\,t}$ is modeled with $d$ length $k$ hidden state vectors
$\vec{s}^{\,d,t}$.  These $d$ states obey a Markov property, as represented in the graph
by an arrow, in that $\vec{s}^{\,i,t}$ only depends on $\vec{s}^{\,i,t-1}$.
In other words,

\begin{equation}
P(S_{k}^{i,t} | \{S\}, \{Y\}; \theta) = P(S_{k}^{i,t} | S_{k}^{i,t-1}; \theta) \quad ,
\end{equation}
where $\{S\}$ and $\{Y\}$ represent the collection of all hidden states and observables
and $\theta$ represents model parameters.
To specify the model, we have a collection of $d$, $k \times k$ transition matrices
(one for each chain) denoted $A^{i}$.  These are directly related to the above
probability.  We also have $d$, $o \times k$ weight matrices $W^{i}$ which
combine linearly with the hidden states to model the observable sequence.
The $o \times 1$ observables are defined as multivariate Gaussian distributed:

\begin{equation}
\vec{y}^{\,t} \sim \mathcal{N}(\sum_{i=1}^{d} W^{i} \cdot \vec{s}^{\,i,t}, C) \quad \quad ,
\end{equation}
where the $d$ matrices $W^{i}$ with shape $o \times k$ are said weights
and $C$ is the $o \times o$ covariance matrix.
Finally, we need to specify $d$ initial state distributions $\vec{\pi}^{\,d}$.
These parameters $W, A, C$, and $\pi$ are all the learnable parameters of the
model, often denoted $\theta$.  The number of independent parameters is
\begin{equation}
{\rm dim} = dok - (d-1)o + d(k-1)k + o^2 + d(k-1) \quad .
\end{equation}
In the first two terms, $dok$ is the number of parameters for $W$, and due to an
overall mean ambiguity  for each $o$ that allows to push the mean into the first
($d=0$) component, we have $dok - (d-1)o$ independent parameters (see canonical $W$ of
the Supplementary Material, Sec.~\ref{appendix:detailed_hessian:preliminary} for
more detail).  The third
term, more transparently written as $dkk - dk$ represents the $dkk$
parameters of $A$ minus the $dk$ probability constraints. The fourth term is the number of
parameters of the covariance matrix. And the fifth and final term, written $dk - d$ is the
number of parameters $dk$ for $\pi$ minus the $d$ probability constraints.

The computational complexities of some key FHMM algorithms are presented in Table~\ref{tab:complexity}.
The exponential scaling of E-step Exact, Log Likelihood, Viterbi, and Hessian algorithms
all stem from the need to evaluate the log likelihood, which contains a sum over
all configurations.  Algorithms such as the Mean Field algorithm instead work to
optimize the Kullback-Leibler divergence (KLD) under a variational approximation,
achieving better complexity.
The mathematical derivations and code implementation details of the FHMM algorithms, including those listed in Table~\ref{tab:complexity}, are presented in the Supplementary Material Secs.~\ref{appendix:fhmm_model_definition}--\ref{appendix:detailed_hessian}.
\arxiv{
\begin{table}[h!]
\centering
\begin{tabular}{ll} \toprule
    \textbf{Algorithm} & \textbf{Complexity}\\ \midrule
    E-step Exact & $\mathcal{O}(Tdk^{d+1})$ \\
    E-step Mean Field & $\mathcal{O}(Tdk^2N_{\rm iter})$ \\
    E-step SVA & $\mathcal{O}(Tdk^2N_{\rm iter})$ \\
    E-step Gibbs & $\mathcal{O}(TdkN_{\rm iter})$ \\ 
    Log Likelihood & $\mathcal{O}(Tdk^{d+1})$ \\
    Viterbi & $\mathcal{O}(Tdk^{d+1})$ \\
    Hessian & $\mathcal{O}(Tdk^{d+1}{\rm dim}^{2})$ \\
    \bottomrule
\end{tabular}
\caption{\label{tab:complexity} Algorithm complexity.  $N_{\rm iter}$ refers to the
         number E-step iterations required per EM iteration.}
\end{table}
}

\peerj{
\begin{table}[h!]
\centering
\begin{tabular}{l|l}
    \textbf{Algorithm} & \textbf{Complexity}\\ \midrule
    E-step Exact & $\mathcal{O}(Tdk^{d+1})$ \\
    E-step Mean Field & $\mathcal{O}(Tdk^2N_{\rm iter})$ \\
    E-step SVA & $\mathcal{O}(Tdk^2N_{\rm iter})$ \\
    E-step Gibbs & $\mathcal{O}(TdkN_{\rm iter})$ \\ 
    Log Likelihood & $\mathcal{O}(Tdk^{d+1})$ \\
    Viterbi & $\mathcal{O}(Tdk^{d+1})$ \\
    Hessian & $\mathcal{O}(Tdk^{d+1}{\rm dim}^{2})$
\end{tabular}
\caption{\label{tab:complexity} {\bf Algorithm complexity}.  $N_{\rm iter}$ refers
         to the number of E-step iterations required per EM iteration.}
\end{table}
}

\subsubsection{Model Selection}
Model selection for FHMMs implies choosing the right number of fluctuators $d$,
and the right number of states $k$, also known as the {\it order selection} problem.
In the following examples, we generally restrict to choosing $d$ and assume $k=2$ (two-level fluctuators).
Reliable determination of the number of hidden fluctuators is a challenging
problem, where many standard methods such as likelihood ratio tests and AIC,
BIC scores can yield poor results ~\cite{Celeux2008}.  Practical methods for
model selection use a variety of model comparison scores in addition to domain
knowledge to help select the appropriate number of valid fluctuators~\cite{Pohle2017}.
Another method, which we pursue here,
is cross-validation. Cross-validation is a generic technique to determine model
performance, using hold-out data and multiple rounds of fitting. It has been shown
to work successfully on hidden Markov models, though it is very computationally
expensive~\cite{Celeux2008}.

To implement robust model selection we combine cross validation, confidence region estimation, and scoring. In effect, we seek to fit the highest $d$ that
confidently generalize across the data. Since for our applications we typically have
plenty of data, the method of cross validation we utilize is to compute the log
likelihood of the model on a hold-out, validation
sequence that immediately follows the training sequence.
We do this for many folds across the whole dataset, to obtain an average estimate.
The log likelihood on the validation sequence is then expected to saturate, or decrease,
beyond the best, most appropriate number of fluctuators $d$.
Confidence interval estimates can aid with the determination of $d$ -- if we don't have
access to confidence intervals, we don't know how trustworthy the individual
point estimates are.
For example, we may fit a $d=5$ model
having a higher log likelihood than $d=4$, but if the fifth fluctuator's weight is commensurate with zero according to the parameter uncertainty, then there is nothing gained over the $d=4$ model.  This emphasizes the utility in having access to confidence interval estimates and will be explored in detail
in later sections.

\subsubsection{Confidence Regions}
We have implemented Hessian-based confidence interval estimation where the Hessian ($H$) is
defined as
\begin{equation}
	H = \frac{\partial^2 \ln \mathcal{L}}{\partial \theta_i \partial \theta_j} \quad ,
\end{equation}
$\mathcal{L}$ is the log likelihood, and $\theta_i$ are the independent parameters
of the model. We then approximate the Observed Information (OI)
matrix as the negative of the Hessian, and take the standard errors of the parameters
as the square root of the inverse of the OI matrix.  For example, if the 000
index element of the $W$ tensor maps to 00 matrix element of the Hessian, we have
the following standard error estimate:
\begin{equation}
dW^{0}_{00} = \sqrt{(-H)^{-1}}_{00}
\end{equation}
The derivation and implementation details of the Hessian calculation
can be found in Sec.~\ref{appendix:detailed_hessian}.

In the context of locally optimal EM fitting of FHMMs in \nomopy \!, a
disadvantage of using the Hessian to compute confidence intervals is that we
assume the fitting procedure has found the globally optimal solution.
However, we may find that the Hessian is giving us high confidence in a
locally optimal solution.  In light of this, we demonstrate the ease of
generating confidence intervals using bootstrapping. Assuming a very
long sequence dataset, we repeatedly estimate the best fit model
to a large number of randomly drawn, shorter subsequences.  The confidence
intervals are then estimated based on the distribution of best fit parameter
values.  These confidence intervals will
generally capture more variability than the Hessian CIs; however, they will also
be more expensive to compute.  This is where \nomopy's built-in parallelization
capability can really shine, easily scaling to high-performance computing (HPC) using Dask \cite{dask2016}.

\subsubsection{Performance}
We incorporate parallelism as well as just-in-time (JIT) compilation to address two major
challenges in fitting FHMMs: (1) successfully finding a global optimum when fitting FHMMs
often requires many attempts, and (2) a number of algorithms suffer from exponential scaling.
When fitting each model we typically perform a number of refits, searching for the
global optimum.
For example, a more complex fitting procedure might be to do an exact method fit on
a schedule of 7 fits, with
5 restarts each, all repeated 10 times, for a total of 350 fits.  This calculation was carried
out for fitting to data generated by four hidden two-level fluctuators on Sandia's HPC
resource Skybridge,
using Dask and \lstinline|dask_jobqueue|~\cite{dask2016}, with each core
operating at 2.6GHz.  To select the most appropriate model, we varied the number of hidden
fluctuators from 2 to 6.  The CPU-hour results are presented in Table~\ref{tab:CPUhours}.
The total carbon footprint of the algorithm is estimated to be about 632 gCO2e,
0.99 kWh, 0.69 tree-months, 3.61km car ride, or 1\% flight from Paris to
London~\cite{Lannelongue2021}.  To put these numbers into perspective, for modeling 1-5 two-level fluctuators, we have a relatively low emission algorithm compared to complex models and simulations, such as weather forecasting and deep learning training, which are in the range $10^{5} - 10^{8}$ gCO2e~\cite{Lannelongue2021}.  We anticipate our algorithm to be in that latter range for $\sim$12 two-level fluctuators.
\arxiv{
\begin{table}
    \centering
    \begin{tabular}{|c|c|c|c|}
        \toprule
        \textbf{Number of chains} & \textbf{Total fit} (minutes) & \textbf{Hessian} (minutes) & \textbf{Carbon footprint} (g CO2e) \\
        \midrule
        2 & 2 & 0.08 & 8.9 \\
        3 & 3 & 0.3 & 13.4 \\
        4 & 17 & 1 & 75.6 \\
        5 & 47 & 3 & 222.4 \\
        6 & 60 & 8 & 302.5 \\
        \bottomrule
    \end{tabular}
    \caption{The performance of FHMM fitting in \nomopy on Sandia National Laboratories' Skybridge HPC cluster using 16 workers with 2 cores each and a total of 350 model fits for each row in the table.}
    \label{tab:CPUhours}
\end{table}
}
\peerj{
\begin{table}
    \centering
    \begin{tabular}{c|c|c|c}
        \textbf{Number of chains} & \textbf{Total fit} (minutes) & \textbf{Hessian} (minutes) & \textbf{Carbon footprint} (g CO2e) \\
        \midrule
        2 & 2 & 0.08 & 8.9 \\
        3 & 3 & 0.3 & 13.4 \\
        4 & 17 & 1 & 75.6 \\
        5 & 47 & 3 & 222.4 \\
        6 & 60 & 8 & 302.5
    \end{tabular}
    \caption{{\bf Performance of the FHMM `exact' fitting algorithm}. The fits
    were done using \nomopy on Sandia National Laboratories' Skybridge HPC cluster using 16 workers with 2 cores each and a total of 350 model fits for each row in the table.}
    \label{tab:CPUhours}
\end{table}
}

\subsection{Noise models}
We include a physically motivated thermal two-level fluctuator (TLF) model within \nomopy \!.  The excitation and relaxation frequencies of the TLF are defined as follows,

\begin{equation}
f_{e/r} = \exp^{(E_b \mp \Delta E / 2) / (k_{\rm B} T)} \quad ,
\end{equation}
for excitation/relaxation ($e/r$) frequencies, where $E_b$ is the barrier energy,
$\Delta E$ is the energy difference between configurations, $k_{\rm B}$
is the Boltzmann constant, and $T$ is the temperature.  Given these frequencies, we construct
the rate matrix

\begin{equation}
M =
\begin{bmatrix}
  -f_e &  f_r \\
   f_e & -f_r
\end{bmatrix} \quad .
\end{equation}
We can then generate noise data using FHMM and the transition matrix $P = e^{M dt}$, where
$dt$ is the sampling period.  The code usage is shown in Sec.~\ref{sec:detailed_noise_models}.

\subsection{Higher order statistics}
One crude measure of the non-Gaussian structure of a time series signal is to histogram
the data and perform a distributional test; however, this method can fail for many types of
non-Gaussian noise. A more sophisticated method of testing for
Gaussianity is to calculate the second spectrum~\cite{Seidler1996, Restle1985, Beck1978}.
Deviation from the Gaussian background second spectrum is then used as a measure of the
non-Gaussianity of the signal. The second spectrum is given by

\begin{equation}
    \langle S_{p}^{(2)}\rangle = 8 T \sum_{k=b_L;n=b_L}^{b_H - p} \langle A_{k+p} A_{k}^{*} A_{n+p}^{*} A_{n} \rangle \quad , 
\end{equation}
where $A_{k}$ are the Fourier transform coefficients of the signal and $b_H$, $b_L$ represent the
band limits.  If the signal is Gaussian, then we have a decoupling resulting in

\begin{equation}
    \langle S_{p}^{(2)}\rangle_{\rm Gaussian} = 8 T \sum_{n=b_L}^{b_H - p} \langle A_{n+p} A_{n+p}^{*}
    A_n A_{n}^{*} \rangle = \frac{2}{T} \sum_{n=b_L}^{b_H - p} \langle S_{n+p}^{(1)} \rangle \langle
    S_{n}^{(1)}\rangle \quad ,
\end{equation}
where $S_{p}^{(1)}$ is the power spectral density. We can also separate the second spectrum into
amplitude and phase components (denoted $S_{p}^{2,a}$ and $S_{p}^{2,\phi}$,
respectively) in order to categorize the source of non-Gaussianity.

We use this analysis to show that while the histogram of a 4 TLF system
statistically tests as Gaussian, a $\chi^2$ test comparing the estimated
second spectrum and the Gaussian background allows us to detect the non-Gaussianity.
The code usage and examples are in Sec.~\ref{sec:detailed_hos}.

From a practical standpoint, one limitation to utilizing the second spectrum is
the apparent need for many time samples to discern non-Gaussianity at lower frequencies.
The analysis typically requires on the order of 10 million samples. 

\section{Detail of capabilities}

In this section, we demonstrate code usage and showcase specific examples applying
\nomopy\!'s capabilities.

\subsection{Factorial Hidden Markov Models}

As described in the Overview, a FHMM is determined by the number of hidden chains $d$, the
number of states for each hidden chain $k$, the number of observable states $o$, and the
length of the time series $T$.  With these parameter definitions,  defining a FHMM in \nomopy
is as follows:
\begin{tcolorbox}
\begin{lstlisting}
from nomopy.fhmm import FHMM
fhmm = FHMM(T=T, d=d, o=o, k=k, em_max_iter=100, method='exact')
\end{lstlisting}
\end{tcolorbox}
\noindent

\subsubsection{Operating modes}

The \lstinline|FHMM| object can be used in a number of different configurations.
In addition to specifying the \lstinline|method| used for EM fitting, we can control
the convergence criteria, stochastic fitting, the number of (random initialization) restarts
to find the best optimum, 
and the number of E step iterations (e.g. for SVA we have \lstinline|sva_max_iter|) and KLD
tolerance if applicable.  Also, we can fix some of the model parameters such that they do
not update during fitting, or fix them all to set the model to a known model.
We can specify initial values for model parameters to use instead of random initialization
when fitting.
These parameter operating modes are described in detail in Section~\ref{sec:codeReference}, Table~\ref{tab:fhmm_parameters}.
We also show the public API methods of the \lstinline|FHMM| class in Section~\ref{sec:codeReference},
Table~\ref{tab:fhmm_methods}.
These functions generally deal with model fit control such as convergence tolerance or
counts; scoring such as calculating the log likelihood; and confidence region estimation
such as Hessian and standard error calculation.

\subsubsection{Fitting data}

Fitting requires that the data, denoted X, have shape \lstinline|(number of samples, T, o)|.
Aside from this specification, the interface mimics the ScikitLearn interface \cite{scikit-learn} though, due
to the multidimensional nature of the problem, it is generally not going to be compatible.
An \lstinline|'exact'| method fit is carried out as follows,
\begin{tcolorbox}
\begin{lstlisting}
fhmm = FHMM(T=T, d=d, o=o, k=k, em_max_iter=100, method='exact')
fhmm.fit(X)
\end{lstlisting}
\end{tcolorbox}
\noindent

\subsubsection{Cross-validation}
\begin{figure}[htp]
	\includegraphics[width=16cm, height=6cm]{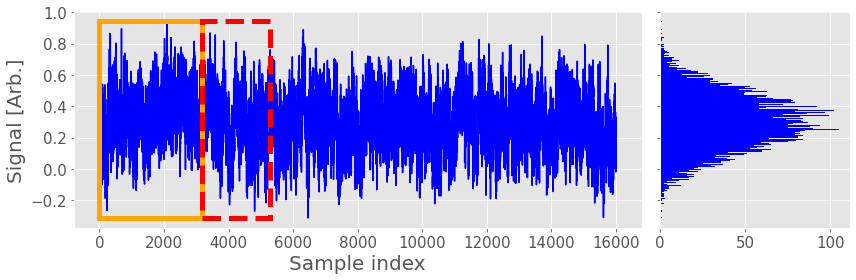}
\caption{\textbf{Cross-validation procedure.} (left) Schematic of the cross-validation folds. The orange solid box is the
         training set, and the red dashed box is the test set. The combined window is
         slid over the data in increments to achieve \lstinline|n_splits| folds.
         (right) A histogram of the time series.}
\label{fig:TSCV}
\end{figure}

Cross-validation is implemented in similar fashion to ScikitLearn where we have \lstinline|FHMMCV|
extending \lstinline|FHMM| to provide built-in (timewise) cross-validation.
\begin{tcolorbox}
\begin{lstlisting}
from nomopy.fhmm import FHMMCV

fhmm = FHMMCV(T=T, d=d, o=o, k=k,
              em_max_iter=100,
              method='exact',
              subsequence_size=0.33,
              test_size=0.4,
              n_splits=20,
              n_jobs=-1)
fhmm.fit(X)
\end{lstlisting}
\end{tcolorbox}
\noindent
We include a schematic of the cross-validation process in Fig.~\ref{fig:TSCV}.  Here, \lstinline|subsequence_size| is about 1/3 of the total data set length and the
test set size (right-most, dashed red box) is about 0.4 or 40\% of that.  The whole window will be slid over the data set in increments resulting in 20 fits.  The additional
parameters and descriptions are in Section~\ref{sec:codeReference},
Table~\ref{tab:fhmmcv_parameters}.  Note, currently this cross-validation method relies on the Viterbi
algorithm to estimate the initial state distribution of the test set.

\subsubsection{Model selection}
Model selection is somewhat of a manual process, but the interface for \lstinline|FHMM|
allows for easy looping:

\begin{tcolorbox}
\begin{lstlisting}
log_likelihoods = []
for d in [1, 2, 3, 4]:
    fhmm = FHMMCV(T=T, d=d, o=o, k=k,
                  em_max_iter=100,
                  method='exact',
                  subsequence_size=0.5,
                  test_size=0.5,
                  n_splits=20,
                  n_jobs=-1)
    fhmm.fit(X)
    log_likelihoods.append(fhmm.log_likelihood())
\end{lstlisting}
\end{tcolorbox}
\noindent
We can then choose the best model based on these validation set log likelihoods (we could have
also looped over \lstinline|k|). For example, we may attempt to quantitatively choose the
model by evaluating the {\it evidence ratio}~\cite{Rudinger2021}:

\begin{equation}
\label{eq:evidenceratio}
e = 2\frac{ \ln{\mathcal{L}_{i}} - \ln{\mathcal{L}_{j}}}{N_i - N_j}
\end{equation} 
where $\ln{\mathcal{L}_{i}}$ is the log likelihood for the $d=i$ model,
$N_i$ is the number of free parameters for the $d=i$ model, and $i$ is the
larger model containing the smaller model $j$.  This
evidence ratio provides (strong) evidence the larger model is better if
$e > 2$.  It provides weak evidence if $1 < e < 2$, and no evidence
if $e < 1$~\cite{Rudinger2021}.

\subsubsection{Bootstrap confidence bounds}
This is also somewhat of a manual process, but essentially we
create a function (here \lstinline|fit_bootstrap_sample|) that samples a random subsequence of the dataset, fits a \lstinline|FHMM|, and saves the result to disk. We can then farm this function out to a cluster using a Dask backend:

\begin{tcolorbox}
\begin{lstlisting}
from joblib import Parallel, delayed, parallel_backend
futures = []
for bs_i in range(number_bootstrap_samples):
    futures.append(delayed(fit_bootstrap_sample)(bs_i))
        
with parallel_backend('dask'):
    res = Parallel()(futures)
\end{lstlisting}
\end{tcolorbox}
\noindent

\subsubsection{Parallelism with Dask}
Scalable parallelism in \nomopy is done with Dask~\cite{dask2016}. The simplest
example of this is calculating the Hessian of the log likelihood.  We launch a
Dask cluster on HPC using \lstinline|SLURMCluster| from \lstinline|dask_jobqueue|, and then
we calculate the Hessian with \lstinline|joblib|'s parallel backend set to use Dask:

\begin{tcolorbox}
\begin{lstlisting}
from joblib import Parallel, delayed, parallel_backend
from dask.distributed import Client
from dask_jobqueue import SLURMCluster
cluster = SLURMCluster(cores=8,
                       processes=4,
                       memory='32GB',
                       project='PROJECT ID',
                       queue='short,batch',
                       job_name='noise',
                       interface='ib0',
                       death_timeout='20s',
                       walltime='04:00:00')
cluster.scale(16)
client = Client(cluster)

with parallel_backend('dask'):
    h = fhmm.hessian()
\end{lstlisting}
\end{tcolorbox}
\noindent

\subsubsection{Examples}

We consider four cases of simulated data analysis, where we vary the number of fluctuators $d$
and the level of white noise $C$.  These are summarized in Table~\ref{tab:paramgrid} for time
steps $T$. The raw time series and histograms are shown in Fig.~\ref{fig:TShists}.  The
power spectral density (PSD) for each case is shown in Fig.~\ref{fig:PSDs}, calculated
using Welch's method (\lstinline|scipy.signal.welch|).  These figures suggest that a limited
amount can be learned about the noise signal from the PSD alone, and only in the case of low
noise can we get an indication from the raw time series of the number of underlying degrees
of freedom. We are able to discover the true underlying model in all cases except for the last row of Table~\ref{tab:paramgrid}.
This last case points to a fundamental limitation of fitting a FHMM in the presence of many
fluctuators and high noise.  We empirically observe that as the noise level increases to
be roughly on the level of the smallest difference between fluctuator weights, fitting fails
more frequently.  However, we are not completely saved by lower noise -- we have also
observed that fitting becomes more challenging with increasing $d$, presumably due to an
increased number of parameters and local minima in the log likelihood landscape.
\arxiv{
\begin{table}[h!]
\centering
\begin{tabular}{ccc} \toprule
    $d$ & $C$ & $T$ \\ \midrule
    2 & 0.0001 & 12800 \\
    2 & 0.01 & 12800 \\
    4 & 0.0001 & 16000 \\
    4 & 0.01 & 16000 \\ \bottomrule
\end{tabular}
\caption{\label{tab:paramgrid} Table of (hyper) parameter experiments.}
\end{table}
}
\peerj{
\begin{table}[h!]
\centering
\begin{tabular}{c|c|c}
    $d$ & $C$ & $T$ \\ \midrule
    2 & 0.0001 & 12800 \\
    2 & 0.01 & 12800 \\
    4 & 0.0001 & 16000 \\
    4 & 0.01 & 16000
\end{tabular}
\caption{\label{tab:paramgrid} \textbf{Experiments.} Each row of this table represents the changed (hyper) parameters for an
example dataset to which we fit a FHMM.}
\end{table}
}
For the $d=2$ ($d=4$) cases, we break up the time series into four samples of length $3200$
($4000$) and perform 20-fold cross-validation over the data, varying the number of fluctuators from 1-4 (1-5).
To find the absolute best fit of model parameters, we do an intensive search on a sample of
size $3200$ ($1000$). We perform this
fitting over different values for $d$ in 1-4 (1-5) in order to score the models using the
evidence ratio (Eq.~\ref{eq:evidenceratio}).  The results are shown in
Figs.~\ref{fig:cved2c0001},~\ref{fig:cved2c01}
(Figs.~\ref{fig:cved4c0001},~\ref{fig:cved4c01}),
where we see an apparent saturation of log likelihood around $d=2$ ($d=4$), and the evidence
ratio suggests the optimal value of $d$.  In the case of $d=4$ and $C=0.01$, we see that
the best model we can fit to the data is $d=3$ -- we seem to be near the limit of the
algorithm's ability to
extract the last fluctuator.  Note, even if the evidence ratio indicated $d=4$,
which can happen,
looking at the fit parameters and their confidence bounds, we are only
able to obtain a confident fit with $d=3$. The weights, white noise level, and log transitions
with their confidence regions, compared to the true values, are shown in
Figs.~\ref{fig:d2c0001W},~\ref{fig:d2c0001A},~\ref{fig:d2c01W},~\ref{fig:d2c01A}
(Figs.~\ref{fig:d4c0001W},~\ref{fig:d4c0001A},~\ref{fig:d4c01W},~\ref{fig:d4c01A}).

\subsection{Noise models}
\label{sec:detailed_noise_models}

In this section we demonstrate how to create thermal TLF noise in \nomopy by
specifying the physical model of the fluctuators (barrier energies, detuning bias energies, temperature, and dipole weights of each fluctuator).

\begin{tcolorbox}
\begin{lstlisting}
from nomopy.noise import ThermalTLFModel

# The physical parameters
d = 4; o=1; k=2
sigma_white_noise = 0.001
w = np.random.rand(d, o, k)               # weights
barrier_energies = [1.1, 0.9, 1.0, 1.2]   # [micro eV]
detuning_energies = [1.5, 0.8, 1.3, 1.0]  # [micro eV]
Temp = 0.12                               # [K]

tlf = ThermalTLFModel(d, sigma_white_noise, dt=1.0)
tlf.set_rates(barrier_energies, detuning_energies, T=Temp)

t, noise = tlf.generate(w, time_steps=10000, n_samples=1, random_seed=1)
\end{lstlisting}
\end{tcolorbox}
\noindent
This will generate a 10k sample time series stored in \lstinline|noise| with
the time values stored in \lstinline|t|.

The \lstinline|ThermalTLFModel| has additional functions for building the rate matrix and
transition matrix, as well as calculating the thermal rates and the analytic
Lorentzian power spectral density (see Section~\ref{sec:codeReference}, Table~\ref{tab:tlf_methods}).

\subsection{Higher order statistics}
\label{sec:detailed_hos}
We have included some simple functions to calculate and work with the second spectrum of
time series data.  

\begin{tcolorbox}
\begin{lstlisting}
from nomopy.hos import second_spectrum

segment_length = len(timeseries) // 300

# Frequency band
fh = 500 # Hz
fl = 100 # Hz

s2, s2_std, s2_gauss, freqs = second_spectrum(timeseries,
                                              dt,
                                              segment_length,
                                              fh,
                                              fl)
\end{lstlisting}
\end{tcolorbox}
\noindent
where \lstinline|s2| will be an array of mean values of the second spectrum calculated
over all segments at frequency values contained in the array \lstinline|freqs|.  The
corresponding array of standard deviations is stored in 
\lstinline|s2_std|, and the Gaussian background is stored in \lstinline|s2_gauss|. To get the phase or amplitude second spectra we must specify the \lstinline|method| as \lstinline|'phase'| or \lstinline|'amplitude'|. Lastly, setting
\lstinline|method='all'| returns all the second spectrum samples:
\begin{tcolorbox}
\begin{lstlisting}
s2s, freqs = second_spectrum(timeseries,
                             dt,
                             segment_length,
                             fh,
                             fl,
                             method='all')
\end{lstlisting}
\end{tcolorbox}
\noindent
This gives an \lstinline|s2s| array of shape \lstinline|(N, len(freqs))|, with \lstinline|N|
being the number of segments (300 above), so we can work
with the distribution of second spectrum values for each frequency.

As a simple statistical test for non-Gaussianity, we use a $\chi^{2}$-test comparing the
second spectrum values of \lstinline|s2| and \lstinline|s2_gauss|.  We first calculate the
errors:

\begin{equation}
    \chi_{i} = \left(\frac{ \langle s_{i}^{(2)}\rangle - \langle s_{i}^{(2)}\rangle_{{\rm Gaussian}}}{\sigma_{i}}\right)^{2}
\end{equation}
where $\sigma_{i}$ is the standard deviation \lstinline|s2_std| / $\sqrt{\rm N}$.
We then compare the sum of errors $\sum_{i} \chi_{i}$ with the $\chi^{2}$ distribution
(by Wilk's theorem) at the 95\% level as a test for non-Gaussianity.  If we can reject
the null hypothesis, then we expect non-Gaussianity.

\subsubsection{Examples}
To showcase the second spectrum analysis, we generate two time series shown in
Fig.~\ref{fig:hosTS}.  The TLF time series was generated using the \lstinline|FHMM| class
of \nomopy\!.  The $1/f^{\beta}$ Gaussian noise with $\beta = 1.2$ was generated using the
algorithm of Timmer and Koenig~\cite{Timmer1995}.  Both signal PSDs display the $1/f$ characteristic, as shown in Fig.~\ref{fig:hosPSD}, and using
a Shapiro-Wilk test, both signal histograms are Gaussian distributions at the 95\% level.
We show the second spectrum for each dataset in Fig.~\ref{fig:hosSecondSpectrum}.
When performing a $\chi^{2}$-test, the 4 TLF system is identified as being non-Gaussian whereas the $1/f^{\beta}$ signal's Gaussian nature is not rejected, both at the 95\% level.

% \begin{figure}[p!]
% \begin{tabular}{c}
% 	\subfloat[d=4 TLF system with noise.  First 100k samples out of 10M points]{\includegraphics[width=16cm, height=6cm]{./Figs/sim_data_time_series_hist.png}} \\
% 	\subfloat[Gaussian $1/f^{\beta}$ noise.  First 100k samples out of 20M points]{\includegraphics[width=16cm, height=6cm]{./Figs/f_data_time_series_hist.png}}
% \end{tabular}
% \caption{Time series traces and histograms used in second spectrum analysis}
% \label{fig:hosTS}
% \end{figure}

% \begin{figure}[p!]
% \begin{tabular}{cc}
% 	\subfloat[d=4 TLF system PSD.]{\includegraphics[width=8cm, height=6cm]{./Figs/sim_data_PSD.png}} &
% 	\subfloat[Gaussian $1/f^{\beta}$ noise PSD.]{\includegraphics[width=8cm, height=6cm]{./Figs/f_data_PSD.png}}
% \end{tabular}
% \caption{Power spectral density of second spectrum example data.}
% \label{fig:hosPSD}
% \end{figure}

% \begin{figure}[p!]
% \begin{tabular}{cc}
% 	\subfloat[d=4 TLF system second spectrum.]{\includegraphics[width=8cm, height=6cm]{./Figs/second_spectrum_d4.png}} &
% 	\subfloat[Gaussian $1/f^{\beta}$ noise second spectrum.]{\includegraphics[width=8cm, height=6cm]{./Figs/second_spectrum_f.png}}
% \end{tabular}
% \caption{Second spectra for the 0.1-0.5 kHz band.  The solid line is the Gaussian
%          background second spectrum. }
% \label{fig:hosSecondSpectrum}
% \end{figure}
\section{Conclusion}

Here we have presented \nomopy \!, a software package written in Python that enables the statistical analysis of time series data as a hidden Markov model or, more generally, a factorial hidden Markov model. The framework of \nomopy includes methods for generating time series assuming a model, parameter estimation based on observed data, evaluating confidence regions for inferred model parameters, and procedures for systematically choosing a model that is most consistent with the data through cross-validation. Our hope is that the simple interface and scalable implementation of \nomopy will allow other researchers to address problems of interest that may have been previously impractical using similar methods.

\section{Acknowledgments}

Sandia National Laboratories is a multi-mission laboratory managed and operated by National Technology and Engineering Solutions of Sandia, LLC., a wholly owned subsidiary of Honeywell International, Inc., for the U.S. Department of Energy's National Nuclear Security Administration under contract DE-NA-0003525.

\arxiv{
\newpage
\section{Code reference}
\label{sec:codeReference}
\begin{longtable}{| l | l | l | }
    \toprule
    \textbf{Parameter} & \textbf{Type, Values} & \textbf{Description} \\ \midrule
	T & int, (Z+) &
        The length of each sequence. \\
    d & int, (Z+) &
        \multicolumn{1}{p{8cm}|}{\raggedright The number of hidden vectors, at each time step.} \\
    k & int, (Z+) &
        \multicolumn{1}{p{8cm}|}{\raggedright The length of each hidden vector, i.e. number of states. } \\
    o & int, (Z+) &
        \multicolumn{1}{p{8cm}|}{\raggedright The length of the output vector. } \\
    n\_restarts & int, (Z+) &
        \multicolumn{1}{p{8cm}|}{\raggedright Number of full model restarts, in search of the global optimum. } \\
    em\_max\_iter & int, (Z+) &
        \multicolumn{1}{p{8cm}|}{\raggedright Maximum number of cycles through E-M steps. } \\
    em\_log\_likelihood\_tol & float, default=1E-8 &
        \multicolumn{1}{p{8cm}|}{\raggedright The tolerance level to discern one log likelihood value from the next. } \\
    em\_log\_likelihood\_count & int, (Z+) &
        \multicolumn{1}{p{8cm}|}{\raggedright Number of log likelihood values without change (according to
        `em\_log\_likelihood\_tol') indicating convergence. } \\
    e\_step\_retries & int, (Z+) &
        \multicolumn{1}{p{8cm}|}{\raggedright Number of random restarts of (applicable) E method. } \\
    method & \multicolumn{1}{p{4cm}|}{\raggedright str, (`gibbs', `mean\_field', `sva', `exact')} &
        \multicolumn{1}{p{8cm}|}{\raggedright Selecting the method for expectation maximization. Options are `gibbs' for Gibbs sampling, `mean\_field' for using mean field estimation (or completely factorized approximation), and `sva' to use the Structured Variational Approximation (SVA), and `exact' for the exact solve (note: very slow for high-dimensional problems).} \\
    gibbs\_max\_iter & int, (Z+) &
        \multicolumn{1}{p{8cm}|}{\raggedright Number of states sampled within Gibbs E-step. } \\
    mean\_field\_max\_iter & int, (Z+) &
        \multicolumn{1}{p{8cm}|}{\raggedright Maximum number of mean field updates.  Once reached, will
        exit without necessarily meeting KLD tolerance. } \\
    mean\_field\_kld\_tol & float, (R+) &
        \multicolumn{1}{p{8cm}|}{\raggedright Tolerance for change in KLD between mean field iterations. } \\
    sva\_max\_iter & int, (Z+) &
        \multicolumn{1}{p{8cm}|}{\raggedright Maximum number of Structured Variational Approximation (SVA) updates.  Once reached, will
        exit without necessarily meeting KLD tolerance. } \\
    sva\_kld\_tol & float, (R+) &
        \multicolumn{1}{p{8cm}|}{\raggedright Tolerance for change in KLD between SVA iterations. } \\
    stochastic\_training & bool &
        \multicolumn{1}{p{8cm}|}{\raggedright Whether or not to use stochastic training -- random and decaying jostling of fit parameters while learning. } \\
    stochastic\_lr & float &
        \multicolumn{1}{p{8cm}|}{\raggedright Roughly the size of the random excursions in fit parameters. } \\
    zero\_probability & float, (R+) &
        \multicolumn{1}{p{8cm}|}{\raggedright Numerical cutoff indicating zero probability (not strictly zero). } \\
    W\_init & numpy.array, None &
        \multicolumn{1}{p{8cm}|}{\raggedright Initialize the starting W weight matrix (shape (d, o, k)), to provide estimation
        a good starting point.  Can be used for debugging or warm starting.  If `None', algorithm will choose
        an initial W. } \\
    A\_init & numpy.array, None &
        \multicolumn{1}{p{8cm}|}{\raggedright Initialize the starting A transition matrix (shape=(d, k, k)), to provide estimation
        a good starting point.  Can be used for debugging or warm starting.  If `None', algorithm will choose
        an initial A. } \\
    C\_init & numpy.array, None &
        \multicolumn{1}{p{8cm}|}{\raggedright Initialize the starting C covariance matrix (shape=(o, o)), to provide estimation
        a good starting point.  Can be used for debugging or warm starting.  If `None', algorithm will choose
        an initial C. } \\
    pi\_init & numpy.array, None &
        \multicolumn{1}{p{8cm}|}{\raggedright Initialize the starting pi initial state distribution matrix (shape=(d, k)), to provide estimation
        a good starting point.  Can be used for debugging or warm starting.  If `None', algorithm will choose
        an initial pi. } \\
    W\_fixed & numpy.array, None &
        \multicolumn{1}{p{8cm}|}{\raggedright Set equal to the true W weight matrix (shape (d, o, k)), to bypass estimation.  Can be
        used for debugging.  If `None', algorithm will update W. } \\
    A\_fixed & numpy.array, None &
        \multicolumn{1}{p{8cm}|}{\raggedright Set equal to the true A transition matrix (shape=(d, k, k)), to bypass estimation.  Can be
        used for debugging.  If `None', algorithm will update A. } \\
    C\_fixed & numpy.array, None &
        \multicolumn{1}{p{8cm}|}{\raggedright Set equal to the true C covariance matrix (shape=(o, o)), to bypass estimation.  Can be
        used for debugging.  If `None', algorithm will update C. } \\
    pi\_fixed & numpy.array, None &
        \multicolumn{1}{p{8cm}|}{\raggedright Set equal to the true pi initial state distribution matrix (shape=(d, k)), to bypass estimation.  Can be
        used for debugging.  If `None', algorithm will update pi. } \\
    verbose & bool, True &
        \multicolumn{1}{p{8cm}|}{\raggedright Print progress and possibly other indicators of algorithm state.} \\ \bottomrule
\caption{\textbf{FHMM parameters.} Parameter descriptions for different operating modes.}
\label{tab:fhmm_parameters}
\end{longtable}
\begin{longtable}{|l|l|}
    \toprule
    \textbf{Method} & \textbf{Description} \\ \midrule
    fhmm.viterbi & \multicolumn{1}{p{8cm}|}{\raggedright Return the viterbi sequence for `sample\_idx'.} \\
    fhmm.fit & \multicolumn{1}{p{8cm}|}{\raggedright Fit the dataset `X' using EM.} \\
    fhmm.hessian & \multicolumn{1}{p{8cm}|}{\raggedright Calculate the Hessian using current model parameters.  Returns Hessian and stores in the class.} \\
    fhmm.standard\_errors & \multicolumn{1}{p{8cm}|}{\raggedright Estimate standard errors based on the Hessian. `hessian()' will be called if needed.
                                                    Returns errors as a tuple $dW, dA, dC, d\pi$ and stores in class. } \\
    fhmm.log\_likelihood & \multicolumn{1}{p{8cm}|}{\raggedright Compute log likelihood of data contained in fhmm.} \\
    fhmm.kld & \multicolumn{1}{p{8cm}|}{\raggedright Kullback-Leibler Divergence, when appropriate for the method selection. } \\
    fhmm.expected\_complete\_log\_likelihood & \multicolumn{1}{p{8cm}|}{\raggedright Expectation value of the complete log likelihood $\langle \ln P \rangle$ . } \\
    fhmm.E & \multicolumn{1}{p{8cm}|}{\raggedright Expectation step.  Returns tuple of hidden state expectations. } \\
    fhmm.M & \multicolumn{1}{p{8cm}|}{\raggedright Maximization step.  Updates model parameters class internally. } \\
    fhmm.is\_fixed & \multicolumn{1}{p{8cm}|}{\raggedright Whether or not model is fixed. } \\
    fhmm.fix\_fit\_params & \multicolumn{1}{p{8cm}|}{\raggedright Fix the model. } \\
    fhmm.unfix\_fit\_params & \multicolumn{1}{p{8cm}|}{\raggedright Unfix the model. } \\
    fhmm.generate & \multicolumn{1}{p{8cm}|}{\raggedright Generates data of specified length
                                             and number of samples based on model parameters.
                                             Optionally return hidden states. } \\
    fhmm.plot\_fit & \multicolumn{1}{p{8cm}|}{\raggedright Convenience plot of data versus Viterbi sequence. } \\
    FHMM.generate\_random\_model\_params & \multicolumn{1}{p{8cm}|}{\raggedright Generates set of model
                                                                    paramters (transition matrices etc.)
                                                                    based on FHMM specification (T, d, etc.).} \\
    \bottomrule
\caption{\textbf{FHMM method API}. Lowercase `fhmm' refers to the class instance, while uppercase
         `FHMM' refers to the class. Function signature is omitted -- see code documentation.}
\label{tab:fhmm_methods}
\end{longtable}
\begin{longtable}{| l | l | l | } \toprule
    \textbf{Parameter} & \textbf{Type, Values} & \textbf{Description} \\ \midrule
    test\_size & float, (0, 1)
        & \multicolumn{1}{p{8cm}|}{\raggedright Test set fraction of subsequence\_size } \\
    subsequence\_size & float, (0, 1)
        & \multicolumn{1}{p{8cm}|}{\raggedright Fraction of total data, giving the length of portions for train/test } \\
    n\_splits & int, (Z+)
        & \multicolumn{1}{p{8cm}|}{\raggedright Number of `subsequence\_size' portions to use for fitting iterations. } \\
    n\_jobs & int, $>$ 0 or -1
        & \multicolumn{1}{p{8cm}|}{\raggedright Number of parallel processes to use for computation, must be
        greater than zero, or equal to -1, indicating to use all
        resources. } \\ \bottomrule
\caption{\textbf{FHMMCV parameters.} Additional parameter descriptions for operating \lstinline|FHMMCV|.}
\label{tab:fhmmcv_parameters}
\end{longtable}
\begin{longtable}{|l|l|}
    \toprule
    \textbf{Method} & \textbf{Description} \\ \midrule
    tlf.set\_rates & \multicolumn{1}{p{8cm}|}{\raggedright Sets model rates from
    physical parameters} \\
    tlf.build\_rate\_matrix & \multicolumn{1}{p{8cm}|}{\raggedright Takes excitation and 
    relaxation frequencies and returns the rate matrix. } \\
    tlf.build\_transition\_matrix & \multicolumn{1}{p{8cm}|}{\raggedright Takes 
    excitation/relaxation frequencies and sample period and returns the rate matrix.} \\
    tlf.calculate\_thermal\_rates & \multicolumn{1}{p{8cm}|}{\raggedright Calculates the
    excitation/relaxation frequencies from model energies and temperature.} \\
    tlf.calculate\_tlf\_psd & \multicolumn{1}{p{8cm}|}{\raggedright Calculates analytic
    Lorentzian PSD.} \\
    \bottomrule
\caption{\textbf{ThermalTLFModel method API.}  Lowercase `tlf' represents an instance of the \lstinline|ThermalTLFModel| class. Function signature is omitted -- see code documentation.}
\label{tab:tlf_methods}
\end{longtable}

}

\peerj{

}

\arxiv{
\appendix
\section{FHMM model definition}
\label{appendix:fhmm_model_definition}

The energy for the model is defined as:

\begin{equation}
\mathcal{H} = \frac{1}{2} \sum_{t=1}^{T} \left[\vec{y}^{\,t}
                - \sum_{i=1}^{d} W^{i} \cdot \vec{s}^{\,i,t} \right]^{\dagger}
              C^{-1} \left[\vec{y}^{\,t}
                - \sum_{i=1}^{d} W^{i} \cdot \vec{s}^{\,i,t} \right]
              - \sum_{t=1}^{T} \sum_{i=1}^{d} \vec{s}^{\,i,t\,\dagger}
                \cdot A^{i} \cdot \vec{s}^{\,i,t-1} \quad ,
\label{eq:energy}
\end{equation}
where the $\dagger$ is used to denote transpose (so as not to confuse with $T$,
for the length of the sequence), and where the $d$ log transition probabilities
$A^{i}$ of shape $k \times k$ are defined and constrained as below,

\begin{equation}
\left[A^{i} \right]_{jl} = \ln P(s_{j}^{\,i,t}|s_{l}^{\,i,t-1}) \quad \quad , \quad \quad
\sum_{j=1}^{k} P(s_{j}^{\,i,t}|s_{l}^{\,i,t-1}) = 1 \quad .
\end{equation}
The $t=1$ case for the second term in equation \ref{eq:energy} is set by the initial
hidden states' probabilities $\vec{\pi}^{\,i}$ such that the second term is:
$\sum_{i=1}^{d} \vec{s}^{\,i,1\,\dagger} \cdot \ln \vec{\pi}^{\,i}$.

The probability model is then defined as follows:
\begin{equation}
P(\{\vec{s}, \vec{y}\}) = \frac{1}{\mathcal{Z}}\,\exp^{-\mathcal{H}(\{\vec{s}, \vec{y}\})}
\quad {\rm where} \quad \mathcal{Z} =
k^{d(T-1)}\left( \frac{(2 \pi)^{o}}{\det C^{-1}} \right)^{T/2}
\end{equation}
where the derivation for the normalization $\mathcal{Z}$ can be found in
Supplementary Section ~\ref{appendix:section:normalization}.

\section{Detailed $\mathcal{Z}$ Derivation}
\label{appendix:section:normalization}

More explicitly, where for simplicity the summation signs are dropped, taking the 
Einstein summation of repeated indices,
\begin{align}
\mathcal{Z} &= \int d^{o}\{y\} \sum_{\{s\}} e^{
               -\frac{1}{2} \left[\vec{y}^{\,t}
                 - W^{i} \cdot \vec{s}^{\,i,t} \right]^{\dagger}
               C^{-1} \left[\vec{y}^{\,t}
                 - W^{j} \cdot \vec{s}^{\,j,t} \right]
               + \vec{s}^{\,i,t\,\dagger}
                 \cdot A^{i} \cdot \vec{s}^{\,i,t-1}} \\
&= \sum_{\{s\}} \left(\int d^{o}\{y\} e^{
   -\frac{1}{2} \left[\vec{y}^{\,t}
     - W^{i} \cdot \vec{s}^{\,i,t} \right]^{\dagger}
   C^{-1} \left[\vec{y}^{\,t}
     - W^{j} \cdot \vec{s}^{\,j,t} \right]} \right)
   e^{\vec{s}^{\,i,t\,\dagger} \cdot A^{i} \cdot \vec{s}^{\,i,t-1}} \\
&= \sum_{\{s\}} \left(\sqrt{\frac{(2 \pi)^{o}}{\det C^{-1}}}\right)^{T}
   e^{\vec{s}^{\,i,t\,\dagger} \cdot A^{i} \cdot \vec{s}^{\,i,t-1}} \\
&= \left(\sqrt{\frac{(2 \pi)^{o}}{\det C^{-1}}}\right)^{T}
   \sum_{\{s\}} e^{\vec{s}^{\,i,t\,\dagger}
     \cdot A^{i} \cdot \vec{s}^{\,i,t-1}} \quad \quad ,
\end{align}
where the reduction comes from $T$ times the usual multivariate Gaussian
normalization, one for each $t$ in the summation within the exponent.
All that is left to show is that the remaining summation piece is equal to
$k^{d(T-1)}$.  First, the $t=1$ term:
\begin{align}
\sum_{\{s\}}\left( \cdots \right)_{t=1}
&= \sum_{\{s\}} e^{\vec{s}^{\,i,1\,\dagger} \cdot \ln \vec{\pi}^{\,i}} \\
&= \sum_{\{s\}} e^{\vec{s}^{\,1,1\,\dagger} \cdot \ln \vec{\pi}^{\,1}} \cdots
   e^{\vec{s}^{\,d,1\,\dagger} \cdot \ln \vec{\pi}^{\,d}} \\
&= \sum_{\vec{s}^{\,1,1}} e^{\vec{s}^{\,1,1\,\dagger} \cdot \ln \vec{\pi}^{\,1}} \cdots
   \sum_{\vec{s}^{\,d,1}} e^{\vec{s}^{\,d,1\,\dagger} \cdot \ln \vec{\pi}^{\,d}} \\
&= \left( \sum_{i_{1}=1}^{k} e^{\ln \pi_{i}^{\,1}}\right)\cdots
   \left( \sum_{i_{d}=1}^{k} e^{\ln \pi_{i_{d}}^{\,d}}\right) \\
&= \left( \sum_{i_{1}=1}^{k} \pi_{i}^{\,1}\right)\cdots
   \left( \sum_{i_{d}=1}^{k} \pi_{i_{d}}^{\,d}\right) \\
&= \left(1\right)\cdots \left(1\right) \\
&= 1 \quad \quad .
\end{align}
And now the rest of the terms:
\begin{align}
\sum_{\{s\}}\left( \cdots \right)_{t>1}
&= \sum_{\{s\}}  e^{\vec{s}^{\,i,t\,\dagger} \cdot A^{i} \cdot \vec{s}^{\,i,t-1}}\\
&= \sum_{\{s\}} e^{\vec{s}^{\,1,2\,\dagger} \cdot A^{1} \cdot \vec{s}^{\,1,1}} \cdots
   e^{\vec{s}^{\,d,T\,\dagger} \cdot A^{d} \cdot \vec{s}^{\,d,T-1}} \\
&= \sum_{\vec{s}^{\,1,2},\vec{s}^{\,1,1}} e^{\vec{s}^{\,1,2\,\dagger} \cdot A^{1} \cdot \vec{s}^{\,1,1}} \cdots
   \sum_{\vec{s}^{\,d,T},\vec{s}^{\,d,T-1}} e^{\vec{s}^{\,d,T\,\dagger} \cdot A^{d} \cdot \vec{s}^{\,d,T-1}} \\
&= \left(\sum_{\vec{s}^{\,1,1}} \sum_{\vec{s}^{\,1,2}} e^{\vec{s}^{\,1,2\,\dagger}
         \cdot A^{1} \cdot \vec{s}^{\,1,1}}\right) \cdots
   \left(\sum_{\vec{s}^{\,d,T-1}} \sum_{\vec{s}^{\,d,T}} e^{\vec{s}^{\,d,T\,\dagger}
         \cdot A^{1} \cdot \vec{s}^{\,d,T-1}}\right) \\
&= \left(\sum_{i_{1}=1}^{k} \sum_{j_{1}=1}^{k} e^{A^{1}_{j_{1} i_{1}}} \right)\cdots
   \left(\sum_{i_{d(T-1)}=1}^{k} \sum_{j_{d(T-1)}=1}^{k} e^{A^{d}_{j_{d(T-1)} i_{d(T-1)}}} \right) \\
&= \left(\sum_{i_{1}=1}^{k} 1\right)\cdots
   \left(\sum_{i_{d(T-1)}=1}^{k} 1\right) \\
&= \left(k\right)\cdots \left(k\right) \\
&= k^{d(T-1)} \quad \quad ,
\end{align}
and we have arrived at the normalization.

\section{Detailed Parameter Estimation}

Minimizing the clamped log probability with respect to the parameters:

\begin{equation}
\mathcal{Q} = \langle -\ln P(\{\vec{s}, \vec{y}\}) \rangle_{\rm c}
= \langle -\mathcal{H} - \ln\mathcal{Z} \rangle_{\rm c}
\end{equation}
First we will solve for the $W^{i}$ matrices via $\partial \mathcal{Q}/\partial W_{jl}^{i} = 0$.
The relevant terms are the $W^{i}$ dependent cross terms and squared term from $\mathcal{H}$:

\begin{align}
\frac{\partial \mathcal{Q}}{\partial W_{jl}^{i}} = &
\frac{1}{2} \sum_{t=1}^{T} s_{l}^{\,i,t} C_{jm}^{-1} y_{m}^{t} \\
&+ \frac{1}{2} \sum_{t=1}^{T} y_{m}^{t} C_{mj}^{-1} s_{l}^{\,i,t} \\
&- \frac{1}{2} \sum_{t=1}^{T} s_{l}^{\,i,t} C_{jm}^{-1}
   \left(\sum_{n}^{d} W_{mp}^{n} s_{p}^{\,n,t}\right) \\
&- \frac{1}{2} \sum_{t=1}^{T} \left(\sum_{n}^{d} W_{mp}^{n} s_{p}^{\,n,t} \right)
   C_{mj}^{-1} s_{l}^{\,i,t}
\end{align}
Since $C^{-1}$ is symmetric, the first two terms combine as well as the last two terms.
We also remove $C^{-1}$ by multiplying through by $C$.

\begin{align}
\frac{\partial \mathcal{Q}}{\partial W_{jl}^{i}} &=
\bigg \langle \sum_{t=1}^{T} s_{l}^{\,i,t} y_{j}^{t}
- \sum_{t=1}^{T} s_{l}^{\,i,t} \left(\sum_{n}^{d} W_{jp}^{n} s_{p}^{\,n,t}\right)\bigg\rangle_{\rm c} \\
&= \sum_{t=1}^{T} \langle s_{l}^{\,i,t}\rangle_{\rm c} y_{j}^{t}
- \sum_{t=1}^{T} \sum_{n}^{d} W_{jp}^{n} \langle s_{p}^{\,n,t}s_{l}^{\,i,t}\rangle_{\rm c} \\
&= \sum_{t=1}^{T} \langle s_{l}^{\,i,t}\rangle_{\rm c} y_{j}^{t}
- \left(\sum_{t=1}^{T} \langle s_{l}^{\,i,t} s_{p}^{\,n,t}\rangle_{\rm c}\right) W_{jp}^{n} \\
&= 0
\end{align}
Taking the combined indices $i$, $l$ and $n$, $p$ and creating stacked vectors/matrices, we
can write this equation as

\begin{equation}
\frac{\partial \mathcal{Q}}{\partial W_{jl}^{i}} =
\sum_{t=1}^{T} \langle \vec{s}^{\,t}\rangle_{\rm c} y_{j}^{t}
- \left(\sum_{t=1}^{T} \langle \vec{s}^{\,t} \vec{s}^{\,t\, \dagger}\rangle_{\rm c}\right) \vec{W}_{j}
\end{equation}
where $\vec{s}^{\,t} \vec{s}^{\,t\, \dagger}$ is the outer-product of two $dk$ dimensional vectors.
This gives a solvable linear system of equations for each $j$, such that

\begin{equation}
W_{A,j} = \left(\sum_{t=1}^{T} \langle \vec{s}^{\,t} \vec{s}^{\,t\, \dagger}\rangle_{\rm c}\right)_{A B}^{-1}
\left[\sum_{t=1}^{T} \langle \vec{s}^{\,t}\rangle_{\rm c} \vec{y}^{\,t} \right]_{B,j}
\end{equation}
where we have combined the indices into $A$ and $B$ such that it is clear we have inverted the
matrix equation for each $j$, corresponding to the $\vec{y}^{\,t}$ components.  The inverse in
this equation is the Moore-Penrose pseudo-inverse. This is the update equation for $W^{i}$.

In order to find the update equation for $A^{i}$ we need to add a Lagrange multiplier for
the probability condition.  Highlighting the important terms regarding applying
$\frac{\partial}{\partial A_{jl}^{i}}$:

\begin{equation}
\mathcal{Q} = \sum_{t=1}^{T} A_{jl}^{i} \langle s_{j}^{i,t} s_{l}^{i,t-1} \rangle_{\rm c}
- \lambda_{il} \left( \sum_{j}^{k} e^{A_{jl}^{i}} - 1\right) + \cdots
\end{equation}

\begin{equation}
\frac{\partial \mathcal{Q}}{\partial A_{jl}^{i}} =
\sum_{t=1}^{T} \langle s_{j}^{i,t} s_{l}^{i,t-1} \rangle_{\rm c} - \lambda_{il} e^{A_{jl}^{i}} = 0
\end{equation}

\begin{equation}
e^{A_{jl}^{i}} = \frac{\sum_{t=1}^{T} \langle s_{j}^{i,t} s_{l}^{i,t-1} \rangle_{\rm c}}{\lambda_{il}}
\end{equation}
The partial derivative $\partial/\partial \lambda_{il}$ enforces the probability constraint

\begin{equation}
\sum_{j=1}^{k}\frac{\sum_{t=1}^{T} \langle s_{j}^{i,t} s_{l}^{i,t-1} \rangle_{\rm c}}{\lambda_{il}} = 1
\end{equation}
We can then finish solving the two equations.

\begin{equation}
\lambda_{il} = \sum_{j=1}^{k} \sum_{t=1}^{T} \langle s_{j}^{i,t} s_{l}^{i,t-1} \rangle_{\rm c}
\end{equation}

\begin{equation}
A_{jl}^{i} = \ln \frac{\sum_{t=1}^{T} \langle s_{j}^{i,t} s_{l}^{i,t-1} \rangle_{\rm c}}
{\sum_{t=1}^{T} \sum_{j=1}^{k} \langle s_{j}^{i,t} s_{l}^{i,t-1} \rangle_{\rm c}}
\end{equation}
This is the update equation for $A^{i}$.  It can be made numerically more stable by expanding
and computing the difference in $\,\ln\,$s.

To compute the estimate for $C$, the relevant terms are

\begin{align}
\mathcal{Q} &= \langle -\mathcal{H} - \ln\mathcal{Z} \rangle_{\rm c} \\
&= \bigg\langle- \frac{1}{2} \sum_{t=1}^{T} \left[\vec{y}^{\,t}
 - W^{q} \cdot \vec{s}^{\,q,t} \right]^{\dagger}
 C^{-1} \left[\vec{y}^{\,t}
 - W^{p} \cdot \vec{s}^{\,p,t} \right] - \ln \mathcal{Z} + \cdots \bigg\rangle_{\rm c} \\
&= \bigg\langle- \frac{1}{2} \sum_{t=1}^{T} \left[\vec{y}^{\,t}
 - W^{q} \cdot \vec{s}^{\,q,t} \right]^{\dagger}
 C^{-1} \left[\vec{y}^{\,t}
 - W^{p} \cdot \vec{s}^{\,p,t} \right] + \frac{T}{2} \ln \det C^{-1}
 + \cdots \bigg\rangle_{\rm c}
\end{align}
First let's look at the matrix derivative of the $\,\ln \det\,$ term.  We have

\begin{equation}
\frac{\partial}{\partial C_{ij}^{-1}} \ln \det C_{ij}^{-1} =
\frac{1}{\det C_{ij}^{-1}} \frac{\partial}{\partial C_{ij}^{-1}} \det C_{ij}^{-1} =
\frac{1}{\det C_{ij}^{-1}} {\rm adj}\,C_{ji}^{-1} = 
C_{ji} = C_{ij}
\end{equation}
We are now ready to calculate the partial derivative:

\begin{equation}
\partial \mathcal{Q}/\partial C_{ij}^{-1} =
\bigg\langle- \frac{1}{2} \sum_{t=1}^{T} \left[y_{i}^{\,t}
 - W_{in}^{q} s_{n}^{\,q,t} \right] \left[y_{j}^{\,t}
 - W_{jm}^{p} s_{m}^{\,p,t} \right]
 + \frac{T}{2} C_{ij}\bigg\rangle_{\rm c}
\end{equation}

\begin{align}
C_{ij} &= \frac{1}{T} \sum_{t=1}^{T}\left[y_{i}^{t} y_{j}^{t} - W_{in}^{q} s_{n}^{q,t} y_{j}^{t}
- y_{i}^{t} W_{jm}^{p} s_{m}^{p,t}
+ W_{in}^{q} W_{jm}^{p} \langle s_{n}^{q,t} s_{m}^{p,t} \rangle_{\rm c}\right]
%&= \frac{1}{T} \sum_{t=1}^{T} \left[y_{i}^{t} y_{j}^{t} -  y_{i}^{t} W_{jm}^{p} s_{m}^{p,t}\right]
\end{align}
where we have used the solution for $W^{i}$ to reduce the equation. This is the update equation
for $C$.

The final parameter to estimate is the initial state distribution, $\pi_{j}^{i}$.  The
relevant terms are the following ($t=1$ from the $A^{i}$ term), where we have added the
probability constraint as a Lagrange multiplier.

\begin{equation}
\mathcal{Q} = \langle s_{j}^{i,1}\rangle_{\rm c} \ln \pi_{j}^{i}
- \lambda_{i} \left(\sum_{j} \pi_{j}^{i} - 1 \right) + \cdots
\end{equation}

\begin{equation}
\frac{\partial\mathcal{Q}}{\partial \pi_{j}^{i}} = \frac{\langle s_{j}^{i,1}\rangle}{\pi_{j}^{i}}
- \lambda_{i}
\end{equation}
From which we have that $\pi_{j}^{i} = \langle s_{j}^{i,1}\rangle_{\rm c} / \lambda_{i}$.  Imposing
the probability condition yields

\begin{equation}
\vec{\pi}^{\,i} = \frac{\langle \vec{s}^{\,i,1}\rangle_{\rm c}}{\sum_{j} \langle s_{j}^{i,1}\rangle_{\rm c}}
\end{equation}
This is the update equation for the remaining parameter $\vec{\pi}^{\,i}$.

\section{Detailed Exact Expectation Estimation}

In order to calculate the hidden state expectations we first make use of the
forward and backward recursion relations.  The forward recursion can be written:
\begin{equation}
    \alpha_{t} = P(Y_t | \{S_t\}) \sum_{\{S_{t-1}\}} \prod_{i=1}^{d} P(S_{t}^{i}|S_{t-1}^{i}) \alpha_{t-1}
\end{equation}
First, we normalize $\alpha$'s in the recurrence relation such that our recurrence relation looks like the
following
(we'll use $\widetilde{\alpha}$ to indicate not yet divided by $c$)
\begin{equation}
\widetilde{\alpha}_{t} = P(Y_t | \{S_t\}) \sum_{\{S_{t-1}\}} \prod_{i=1}^{d} P(S_{t}^{i}|S_{t-1}^{i}) \widehat{\alpha}_{t-1}
\label{eq:recursion}
\end{equation}

\begin{equation}
\widehat{\alpha}_{t-1} = \widetilde{\alpha}_{t-1} / c_{t-1}
\end{equation}
where $c_{t-1} = \sum_{\{S_{t-1}\}} \widetilde{\alpha}_{t-1}$.
In the above, $\widetilde{\alpha}_{t}$ can be thought of as a function of possible $S_{t}^{i}$ (binary) values.
Or, when programming, a vector of length $d^{k}$ with entries containing an evaluation of
$\widetilde{\alpha}_{t}$ for each configuration of $S_{t}$.
Calculating the forward relation with this normalization makes the numerical routine more stable and
also allows for an easy method to track the $c$'s and calculate the likelihood.
\begin{align}
\widetilde{\alpha}_{T} &= P(Y_T | \{S_T\}) \sum_{\{S_{T-1}\}} \prod_{i=1}^{d} P(S_{T}^{i}|S_{T-1}^{i}) \widehat{\alpha}_{T-1} \\
    &= \left(\prod_{j=1}^{T-1} \frac{1}{c_{j}}\right)P(Y_T | \{S_T\}) \sum_{\{S_{T-1}\}} \prod_{i=1}^{d} P(S_{T}^{i}|S_{T-1}^{i}) \alpha_{T-1}
\end{align}
Now when we sum over all hidden states we get
\begin{equation}
c_{T} = \left(\prod_{j=1}^{T-1} \frac{1}{c_{j}}\right) \sum_{\{S_{T}\}} \alpha_{T} \
        \quad \rightarrow \quad \prod_{j=1}^{T} c_{j} = P(\{Y\}| \phi) \;.
\end{equation}
This yields our final relation for the log likelihood:
\begin{equation}
\ln \mathcal{L} = \ln P(\{Y\}| \phi) = \sum_{j=1}^{T} \ln c_{j} \;.
\label{eq:ll}
\end{equation}
To implement the forward recursion, we first initialize the \lstinline|alpha| by
filling the \lstinline|t=0| element as the hidden state realization probability
times the observable probability, $\prod_{i=1}^{d} \pi^{i} P(Y_1 | \{S_1\})$, :
\begin{tcolorbox}
\begin{lstlisting}
alpha = np.ones(shape=(self.T, self.k**self.d))
for i in range(realizations.shape[1]):
    pi = 1
    for d in range(self.d):
        pi *= self.pi[d, realizations[d, i]]

    alpha[0, i] = pi * py[0, i] + eps
\end{lstlisting}
\end{tcolorbox}
\noindent
We can then carry out the recursion relations via
\begin{tcolorbox}
\begin{lstlisting}
c[0] = alpha[0, :].sum()
alpha[0, :] /= c[0]

for t in range(1, self.T):
    for j in range(realizations.shape[1]):
        prob_j = 1
        for d in range(self.d):
            prob_j *= np.exp(self.A)[d,
                                     realizations[d, j],
                                     realizations[d, :]]
        alpha[t, j] = np.sum(alpha[t-1] * prob_j * py[t, j])
    c[t] = alpha[t, :].sum()
    alpha[t, :] /= c[t]
\end{lstlisting}
\end{tcolorbox}
\noindent
The inner \lstinline|for| loop calculates the transition probability product,
$\prod_{i=1}^{d} P(S_{t}^{i}|S_{t-1}^{i})$, keeping
realization of $S_{t-1}$ as an index for summation indicated by \lstinline|:|.
The element of \lstinline|alpha[t, j]| is
then filled out by summing over the realization index element-wise multiplication of
\lstinline|alpha[t-1, :]| and our
precomputed $P(Y_t | \{S_t\})$, \lstinline|py[t, j]|.
We subsequently normalize \lstinline|alpha| and store the normalization.  This
normalization is used to compute the log likelihood, as previously derived.

There is also the related backward recursion relation, namely:
\begin{equation}
\beta_{t-1} = \sum_{\{S_{t}\}} \prod_{i=1}^{d} P(S_{t}^{i}|S_{t-1}^{i}) P(Y_t | \{S_t\}) \beta_{t}
\end{equation}
Using notation similar to the forward ralation above we can write:
\begin{equation}
  \widetilde{\beta}_{t-1} = \sum_{\{S_{t}\}} \prod_{i=1}^{d} P(S_{t}^{i}|S_{t-1}^{i}) P(Y_t | \{S_t\}) \widehat{\beta}_{t}
\end{equation}

\begin{equation}
    \widehat{\beta}_{t-1} = \widetilde{\beta}_{t-1} / c_{t-1}
\end{equation}
where $c_{t-1} = \sum_{\{S_{t-1}\}} \widetilde{\beta}_{t-1}$.
We initialize the \lstinline|beta| as an array with shape
$(T, {\rm number\; of\; realizations})$ and normalization \lstinline|c| as follows:
\begin{tcolorbox}
\begin{lstlisting}
beta = np.ones(shape=(self.T, self.k**self.d))
c[self.T-1] = beta[self.T-1, :].sum()
beta[self.T-1, :] /= c[self.T-1]
\end{lstlisting}
\end{tcolorbox}
\noindent
We then proceed to implement the recursion as follows:
\begin{tcolorbox}
\begin{lstlisting}
for t in reversed(range(1, self.T)):
    for j in range(0, realizations.shape[1]):
        prob_j = 1
        for d in range(self.d):
            prob_j *= np.exp(self.A)[d,
                                     realizations[d, :],
                                     realizations[d, j]]
        beta[t-1, j] = np.sum(beta[t] * prob_j * py[t, :])
    c[t-1] = beta[t-1, :].sum()
    beta[t-1, :] /= c[t-1]
\end{lstlisting}
\end{tcolorbox}
\noindent
The inner \lstinline|for| loop calculates the transition probability product,
$\prod_{i=1}^{d} P(S_{t}^{i}|S_{t-1}^{i})$, keeping
realization of $S_{t}$ as an index for summation.  The element of \lstinline|beta[t-1, j]| is
then filled out by summing over the realization index element-wise multiplication of
\lstinline|beta[t, :]| and our
precomputed $P(Y_t | \{S_t\})$, \lstinline|py[t, :]|.
We subsequently normalize \lstinline|beta| and store the normalization.

We are now ready to calculate the state expectations $\langle S_{t}^{i} \rangle$,
$\langle S_{t}^{i} S_{t}^{j} \rangle$, and $\langle S_{t-1}^{i} S_{t}^{i} \rangle$,
and we do this making use of the following (using shorthand where the absence of an
index means all of them are present: $Y \rightarrow Y_1,...,Y_T$, and
$S_{t} \rightarrow S_{t}^{1}, ..., S_{t}^{d}$):
\begin{equation}
\gamma = P(S_{t} | Y) = \frac{P(S_{t}, Y)}{P(Y)}
\end{equation}
where, due to the dependency graph,
\begin{equation}
\alpha_t \beta_t = P(S_t, Y_1, ...Y_t) P(Y_{t}, ..., Y_{T} | S_{t}) = P(S_{t}, Y)
\end{equation}
and, noting that $P(Y) = \sum_{S_t} P(S_t, Y)$, we have
\begin{equation}
\gamma = P(S_{t} | Y) = \frac{\alpha_t \beta_t}{\sum_{S_t} \alpha_t \beta_t}
       = \frac{\widehat{\alpha}_t \widehat{\beta}_t}{\sum_{S_t} \widehat{\alpha}_t \widehat{\beta}_t}
\end{equation}
where the equality for the hatted case is due to the normalization being multiplicatively
factored out of the numerator and denominator, canceling.
The expectation value of $S_{t}^{i}$ is written as
\begin{equation}
\langle S_{t}^{i} \rangle = \sum_{S_t} S_{t}^{i} P(S_t | Y) \,\, .
\end{equation}
Now, since a particular $S_{t}^{i}$ is either 0 or 1, we can ignore its contribution to the
sum, and replace $S_{t}^{i}$ with 1, yielding
\begin{equation}
\langle S_{t}^{i} \rangle = \sum_{\{\widehat{S_t^{i}}\}} P(S_t | Y) =  \sum_{\{\widehat{S_t^{i}}\}} \gamma_t
\end{equation}
where we use a hat to denote summation over all except the hatted value.

The calculation of $\gamma$ and the state expectation is simply implemented using our
\lstinline|alpha[t]| and \lstinline|beta[t]|.  We first calculate \lstinline|gamma|:
\begin{tcolorbox}
\begin{lstlisting}
gamma = alpha * beta
norm = gamma.sum(axis=1, keepdims=True)
gamma /= norm
\end{lstlisting}
\end{tcolorbox}
\noindent
where the norm is a summation over the realization index. We then proceed to implement
the expectation calculation very simply as a bunch of \lstinline|for| loops:

\begin{tcolorbox}
\begin{lstlisting}
s_exp = eps * np.ones(shape=(self.T, self.d, self.k))
for t in range(self.T):
    for d in range(self.d):
        for k in range(self.k):
            indices = list(k_contrib[(d, k)])
            if indices:
                s_exp[t, d, k] += np.sum(gamma[t, indices])
\end{lstlisting}
\end{tcolorbox}
\noindent
We initialize an empty matrix to hold the expectation and loop over all indices.
In the inner-most \lstinline|for| loop, we grab all of the realizations where this
particular \lstinline|d| and \lstinline|k| are 1 (equivalent to setting $S_{t}^{i}$
to 1 as described above) and we sum over the selection of only these realizations,
\lstinline|gamma[t, indices]|.

The calculation of \(\langle S_{t}^{i} S_{t}^{j} \rangle\) is done in a similar
manner to above. The expectation value is written as
\begin{equation}
\langle S_{t}^{i} S_{t}^{j} \rangle = \sum_{S_t} S_{t}^{i} S_{t}^{j} P(S_t | Y) \,\, .
\end{equation}
Now, since the product $S_{t}^{i} S_{t}^{j}$ is only 1 when both are 1, otherwise 0, we
can alter the summation and replace $S_{t}^{i} S_{t}^{j}$ with 1, yielding
\begin{equation}
\langle S_{t}^{i} S_{t}^{j} \rangle = \sum_{\{\widehat{S_t^{i}},\widehat{S_t^{j}}\}} P(S_t | Y)
                                    = \sum_{\{\widehat{S_t^{i}},\widehat{S_t^{j}}\}} \gamma_t
\end{equation}
where we use a hat to denote summation over all except the hatted values.  We
implement this in code by looping over all index values of the expectation:
\begin{tcolorbox}
\begin{lstlisting}
ss_exp = eps * np.ones(shape=(self.T, self.d, self.d, self.k, self.k))
for t in range(self.T):
    for d1 in range(self.d):
        for d2 in range(self.d):
            for k1 in range(self.k):
                for k2 in range(self.k):
                    indices = list(k_contrib[(d1, k1)] & k_contrib[(d2, k2)])
                    if indices:
                        ss_exp[t, d1, d2, k1, k2] += np.sum(gamma[t, indices])
\end{lstlisting}
\end{tcolorbox}
\noindent
In the inner-most \lstinline|for| loop we restrict to realizations (\lstinline|indices|)
that have the $i$ and $j$, here \lstinline|k1| and \lstinline|k2|, state as 1.  We
assign the expectation element by summing over only these realizations.

The calculation of $\langle S_{t-1}^{i} S_{t}^{i} \rangle$ is a bit trickier.
\begin{equation}
\langle S_{t-1}^{i} S_{t}^{i} \rangle = \sum_{S_{t-1} S_t} S_{t-1}^{i} S_{t}^{i} P(S_{t-1}, S_t | Y) \,\, .
\end{equation}
Similar to above we can replace this sum with

\begin{equation}
\langle S_{t-1}^{i} S_{t}^{i} \rangle = \sum_{\{\widehat{S_{t-1}^{i}},\widehat{S_t^{i}}\}} P(S_{t-1}, S_t | Y) \,\, .
\end{equation}
Now since $P(S_{t-1}, S_t | Y) = P(S_{t-1}, S_t, Y) / P(Y)$ and
$P(Y) = \sum_{S_{t-1}, S_t} P(S_{t-1}, S_t, Y)$, we can use the relation

\begin{equation}
P(S_{t-1}, S_{t}, Y) = \alpha_{t-1} \prod_{i=1}^{d} P(S_t^{i}|S_{t-1}^{i}) P(Y_t|S_{t}) \beta_{t}
\end{equation}
to find that

\begin{equation}
\langle S_{t-1}^{i} S_{t}^{i} \rangle = \frac{\sum_{\{\widehat{S_{t-1}^{i}},\widehat{S_t^{i}}\}}
    \alpha_{t-1} \prod_{i=1}^{d} P(S_t^{i}|S_{t-1}^{i}) P(Y_t|S_{t}) \beta_{t}}
    {\sum_{S_{t-1} S_t} \alpha_{t-1} \prod_{i=1}^{d} P(S_t^{i}|S_{t-1}^{i}) P(Y_t|S_{t}) \beta_{t}} \,\, .
\end{equation}
In order to implement this in code we first store the values of all possible values of the
product of transition probabilities, $\prod_{i=1}^{d} P(S_t^{i}|S_{t-1}^{i})$,
which we call \lstinline|psstm1|:

\begin{tcolorbox}
\begin{lstlisting}
psstm1 = np.ones(shape=(realizations.shape[1],
                         realizations.shape[1]))
for t_i in range(realizations.shape[1]):
    for tm1_j in range(realizations.shape[1]):
        for d in range(self.d):
            psstm1[t_i, tm1_j] *= \
                    np.exp(self.A)[d,
                                   realizations[d, t_i],
                                   realizations[d, tm1_j]]
\end{lstlisting}
\end{tcolorbox}
\noindent
So all we need to do is specify the $t-1$ and $t$ realization indices at obtain the
transition probability product value.  With this in hand we loop over all possible
indices of the expectation value:

\begin{tcolorbox}
\begin{lstlisting}
sstm1_exp = eps * np.ones(shape=(self.T, self.d, self.k, self.k))
for t in range(1, self.T):
norm_t = eps
for d in range(self.d):
    for k1 in range(self.k):
        for k2 in range(self.k):
            t_indices = list(k_contrib[(d, k1)])
            tm1_indices = list(k_contrib[(d, k2)])

            comb_indices = np.transpose([np.repeat(t_indices, len(tm1_indices)),
                                         np.tile(tm1_indices, len(t_indices))])
            comb_t_indices = comb_indices[:, 0]
            comb_tm1_indices = comb_indices[:, 1]

            sstm1_exp[t, d, k1, k2] += np.sum(alpha[t-1, comb_tm1_indices]
                                              * psstm1[comb_t_indices,
                                                       comb_tm1_indices]
                                              * py[t, comb_t_indices]
                                              * beta[t, comb_t_indices])
        
            # Running sum for normalization
            norm_t += sstm1_exp[t, d, k1, k2]

sstm1_exp[t, :, :, :] /= (norm_t/self.d)
\end{lstlisting}
\end{tcolorbox}
\noindent
In the inner-most \lstinline|for| loop, we extract the realizations that have the
\lstinline|k1| and \lstinline|k2| states set to 1, stored in \lstinline|t_indices|
and \lstinline|tm1_indices|, respectively.  Then we form the indices'
cartesian product, using \lstinline|np.repeat| and \lstinline|np.tile| to effect
the loop over all pairs of realizations in a vectorized manner. Then, we fill out
the expectation element summing over this restricted set of indices.  Finally,
we add the value to a running sum for the denominator of the expectation value.
Since we add over all $d$, we overcount the normalization $d$ times. In the last
line of the code snippet, we normalize expectation at the end of each $t$ iteration.

With these exact computations of $\langle S_{t}^{i} \rangle$,
$\langle S_{t}^{i} S_{t}^{j} \rangle$, and $\langle S_{t-1}^{i} S_{t}^{i} \rangle$,
the Expectation-Maximization algorithm can continue on to the Maximization step.

\section{Detailed Mean Field Expectation Estimation}

Here we are only going to derive the Mean Field (MF) estimation of expectation values.  We start
with the MF Hamiltonian:

\begin{equation}
\mathcal{H}_{\rm MF} = \frac{1}{2} \sum_{t=1}^{T}
\left[\vec{y}^{\,t} - \vec{\mu}^{\,t}\right]^{\dagger}
C^{-1}
\left[\vec{y}^{\,t} - \vec{\mu}^{\,t}\right]
- \sum_{t=1}^{T} \sum_{i=1}^{d}
\vec{s}^{\,i,t\,\dagger} \cdot \ln \vec{m}^{\,i,t}\quad ,
\end{equation}
thus the probability completely factorizes,

\begin{equation}
\widetilde{P}(\{\vec{s},\vec{y}\}) = \frac{1}{\mathcal{Z}_{\rm MF}}
\prod_{t}^{T} e^{\left[\vec{y}^{\,t} - \vec{\mu}^{\,t}\right]^{\dagger}
C^{-1}
\left[\vec{y}^{\,t} - \vec{\mu}^{\,t}\right]}
\prod_{t,i,j}^{T,d,k}
\left(m_{j}^{i,t}\right)^{s_{j}^{i,t}}
\quad {\rm where} \quad \mathcal{Z}_{\rm MF} =
\left( \frac{(2 \pi)^{o}}{\det C^{-1}} \right)^{T/2}
\end{equation}
The normalization factor is that for the multivariate Gaussian, as before. There is no
contribution from the product factor of $m_{j}^{i,t}$ as we now explain.
Each product of $m_{j}^{i,t}$ is a multinomial distribution, for each $i,t$, provided
$\sum_{j} m_{j}^{i,t} = 1$.
The multinomial distribution is defined as choosing $n$ states out of $k$ possible
states.  With $x_j$ as the number of states $j$ chosen, the probability mass function is

\begin{equation}
P(X_{1} = x_{1}, ..., X_{k} = x_{k}) = \frac{n!}{x_{1}!\cdots x_{k}!} p_{1}^{x_{1}} \cdots p_{k}^{x_{k}}
\quad {\rm where} \quad \sum_{j}^{k} x_{j} = n
\end{equation}
In our situation, the number of states chosen is $n=1$, and $k$ is still conveniently the number
of states. The number of states for $j$ is replaced with our vector $s_{j}^{i,t}$, where we
abuse notation slightly denoting the
random state vector as capital $S_{j}^{i,t}$.  And finally, the probabilities $p_{j}$ are
replaced with $m_{j}^{i,t}$. In which case, suppressing the $i,t$ indices, the PMF becomes

\begin{equation}
P(S_{1} = s_{1}, ..., S_{k} = s_{k}) = \frac{n!}{s_{1}!\cdots s_{k}!} m_{1}^{s_{1}} \cdots m_{k}^{s_{k}}
\quad {\rm where} \quad \sum_{j}^{k} s_{j} = 1
\end{equation}
and since $n=1$ and only one of the $s_{i}$ can be $1$ with all of the others $0$, the
normalization is $1$, yielding,

\begin{equation}
P(S_{1} = s_{1}, ..., S_{k} = s_{k}) = m_{1}^{s_{1}} \cdots m_{k}^{s_{k}}
\quad {\rm where} \quad \sum_{j} s_{j} = 1
\end{equation}
and in particular, reintroducing the $i,t$ indices,
$P(S_{1}^{i,t} = 1, ..., S_{k}^{i,t} = 0) = m_{1}^{i,t}$, for each $i,t$.

On a final note about our case of the $n=1$ multinomial distribution, as will be used later,
$\langle s_{j} \rangle = m_{j}$ and $\langle s_{j} s_{j} \rangle = m_{j}$.

\begin{align}
\langle s_{j} \rangle &= \sum_{s_{j}} s_{j} P(s_{j}) \\
&= m_{j} \frac{\partial}{\partial m_{j}} \sum_{s_{j}} P(s_{j}) \quad \quad &{\rm (N.S.)} \\
&= m_{j} \frac{\partial}{\partial m_{j}} (m_{1} + ... + m_{k}) \quad \quad &{\rm (N.S.)} \\
&= m_{j} \quad \quad .
\end{align}
Similarly,
\begin{align}
\langle s_{j}^{2} \rangle &= \sum_{s_{j}} s_{j}^{2} P(s_{j}) \\
&= m_{j} \frac{\partial}{\partial m_{j}} m_{j} \frac{\partial}{\partial m_{j}}
\sum_{s_{j}} P(s_{j}) \quad \quad &{\rm (N.S.)} \\
&= m_{j} \frac{\partial}{\partial m_{j}} m_{j} \quad \quad &{\rm (N.S.)} \\
&= m_{j} \quad \quad .
\end{align}
It is also now clear that $\langle s_{i} s_{j} \rangle = 0$ for $i \ne j$.
Due to the complete factorization of the distribution function we have the sets of equations:

\begin{equation}
  \langle s_{j}^{i,t} s_{l}^{n,t}\rangle =
  \begin{cases}
    m_{j}^{i,t} m_{l}^{n,t} &i \ne n \\
    m_{j}^{i,t} \delta_{jl} &i = n
  \end{cases}
\end{equation}
We implement this expectation value as follows:
\begin{tcolorbox}
\begin{lstlisting}
mm = np.zeros(shape=(self.T, self.d, self.k, self.d, self.k))
for t in range(self.T):
    mm[t] = np.outer(m[t].ravel(), m[t].ravel())
    mm[t] = mm[t].reshape(self.d, self.k, self.d, self.k)

# Fix diagonal d1 = d2 case:
for t in range(self.T):
    for d in range(self.d):
        mm[t, d, :, d, :] = np.diag(m[t, d, :])

ss_exp = np.swapaxes(mm, 2, 3)
\end{lstlisting}
\end{tcolorbox}
\noindent
where we first calculate for (each $t$) the $(d k, d k)$ outer product
of \lstinline|m[t]|, after first unraveling each \lstinline|m[t]| (of
shape $(d, k)$) into
shape $(dk)$, where the unraveling of the last index is the fastest,
and the first index is slowest.  We then reshape this outer product
into an array of
shape $(d, k, d, k)$, taking care that the index ordering is preserved,
where the last index is changing fastest, up to the first index being
slowest.
After handling the diagonal special case, we then swap the middle $k$
and $d$ to save in our conventional format \lstinline|ss_exp| of shape
$(t, d_1, d_2, k_1, k_2)$.

The final expectation value to calculate is the same-chain, across-time
expecation value:
\begin{align}
\langle s_{j}^{i,t} s_{l}^{i,t-1} \rangle
= \langle s_{j}^{i,t}\rangle\langle s_{l}^{i,t-1} \rangle
= m_{j}^{i,t} m_{l}^{i,t-1}
\end{align}
which we implement in code as follows:
\begin{tcolorbox}
\begin{lstlisting}
sstm1_exp = np.zeros(shape=(self.T, self.d, self.k, self.k))

for t in range(1, self.T):
    for d in range(self.d):
        sstm1_exp[t, d, :, :] = np.outer(m[t, d, :], m[t-1, d, :])
\end{lstlisting}
\end{tcolorbox}
\noindent
where we are using our conventional format for \lstinline|sstm1_exp| of
shape $(t, d, k_1, k_2)$, and we skip the index \lstinline|t=0| since there
is no \lstinline|t=-1| state.

We have shown the expressions and calculations for the expectation values
present in the parameter estimation equations, in terms of
\lstinline|m[t, d, k]|.  Now all that is left to
estimate is the mean field parameter $\vec{m}^{\,i,t}$. It
is estimated via minimizing the Kullback-Leibler divergence (KLD) between
the model distribution and the MF distribution.  The KLD is defined
as

\begin{equation}
\mathcal{KL} = \langle \ln \widetilde{P} \rangle_{\widetilde{P}}
 - \langle \ln P \rangle_{\widetilde{P}}
\end{equation}
Using the definition already covered for $P$ and $\widetilde{P}$, we have:

\begin{align}
\mathcal{KL} = &
\sum_{t,i,j}^{T,d,k} \langle s_{j}^{i,t} \rangle_{\widetilde{P}} \ln m_{j}^{i,t}
- \ln \mathcal{Z}_{\rm MF} \\
&+ \frac{1}{2} \sum_{t=1}^{T} y_{i}^{t} C_{ij}^{-1} y_{j}^{t}
- \frac{1}{2} \sum_{t=1}^{T} W_{nl}^{i} \langle s_{l}^{i,t}\rangle_{\widetilde{P}} C_{np}^{-1} y_{p}^{t}
- \frac{1}{2} \sum_{t=1}^{T} y_{p}^{t} C_{pn}^{-1} W_{nl}^{i} \langle s_{l}^{i,t}\rangle_{\widetilde{P}} \\
&+ \frac{1}{2} \sum_{t=1}^{T} W_{nl}^{i} C_{np}^{-1} W_{pq}^{j}
  \langle s_{l}^{i,t} s_{q}^{j,t} \rangle_{\widetilde{P}}
- \sum_{t=2}^{T} A_{jl}^{i} \langle s_{j}^{i,t} s_{l}^{i,t-1} \rangle_{\widetilde{P}}
- \langle s_{j}^{i,1} \rangle_{\widetilde{P}} \ln \pi_{j}^{i}
+ \ln \mathcal{Z} \\
= &
\sum_{t,i,j}^{T,d,k} \langle s_{j}^{i,t} \rangle_{\widetilde{P}} \ln m_{j}^{i,t}
- \ln \mathcal{Z}_{\rm MF}
+ \frac{1}{2} \sum_{t=1}^{T} y_{i}^{t} C_{ij}^{-1} y_{j}^{t}
- \sum_{t=1}^{T} y_{p}^{t} C_{pn}^{-1} W_{nl}^{i} \langle s_{l}^{i,t}\rangle_{\widetilde{P}} \\
&+ \frac{1}{2} \sum_{t=1}^{T} W_{nl}^{i} C_{np}^{-1} W_{pq}^{j}
  \langle s_{l}^{i,t} s_{q}^{j,t} \rangle_{\widetilde{P}}
- \sum_{t=2}^{T} A_{jl}^{i} \langle s_{j}^{i,t} s_{l}^{i,t-1} \rangle_{\widetilde{P}}
- \langle s_{j}^{i,1} \rangle_{\widetilde{P}} \ln \pi_{j}^{i}
+ \ln \mathcal{Z} \\
= &
\sum_{t,i,j}^{T,d,k} m_{j}^{i,t} \ln m_{j}^{i,t}
- \ln \mathcal{Z}_{\rm MF}
+ \frac{1}{2} \sum_{t=1}^{T} y_{i}^{t} C_{ij}^{-1} y_{j}^{t}
- \sum_{t=1}^{T} y_{p}^{t} C_{pn}^{-1} W_{nl}^{i} m_{l}^{i,t} \\
&+ \frac{1}{2} \sum_{\substack{t=1 \\ i \ne j}}^{T} W_{nl}^{i} C_{np}^{-1} W_{pq}^{j}
  m_{l}^{i,t} m_{q}^{j,t}
+ \frac{1}{2} \sum_{t=1}^{T} W_{nl}^{i} C_{np}^{-1} W_{pq}^{i} m_{l}^{i,t} \delta_{lq} \\
&- \sum_{t=2}^{T} A_{jl}^{i} m_{j}^{i,t} m_{l}^{i,t-1}
- m_{j}^{i,1} \ln \pi_{j}^{i}
+ \ln \mathcal{Z} \\
= &
\sum_{t,i,j}^{T,d,k} m_{j}^{i,t} \ln m_{j}^{i,t}
- \ln \mathcal{Z}_{\rm MF}
+ \frac{1}{2} \sum_{t=1}^{T} y_{i}^{t} C_{ij}^{-1} y_{j}^{t}
- \sum_{t=1}^{T} y_{p}^{t} C_{pn}^{-1} W_{nl}^{i} m_{l}^{i,t} \\
& - \frac{1}{2} \sum_{t=1}^{T} W_{nl}^{i} C_{np}^{-1} W_{pq}^{i} m_{l}^{i,t} m_{q}^{i,t}
+ \frac{1}{2} \sum_{t=1}^{T} W_{nl}^{i} C_{np}^{-1} W_{pq}^{j} m_{l}^{i,t} m_{q}^{j,t}
+ \frac{1}{2} \sum_{t=1}^{T} W_{nl}^{i} C_{np}^{-1} W_{pq}^{i} m_{l}^{i,t} \delta_{lq} \\
&- \sum_{t=2}^{T} A_{jl}^{i} m_{j}^{i,t} m_{l}^{i,t-1}
- m_{j}^{i,1} \ln \pi_{j}^{i}
+ \ln \mathcal{Z} + \lambda_{i,t} \left(\sum_{j=1}^{k} m_{j}^{i,t} - 1 \right)
\end{align}
where we have added the Lagrange multiplier for the probability constraint in the final equality.
We are now ready to estimate the $\vec{m}^{\,i}$ that will minimize the KLD.

\begin{align}
\frac{\partial\mathcal{KL}}{\partial m_{j}^{i,t>1}} = &
\ln m_{j}^{i,t} + 1
- y_{p}^{t} C_{pn}^{-1} W_{nj}^{i}
- \frac{1}{2} W_{nj}^{i} C_{np}^{-1} W_{pq}^{i} m_{q}^{i,t}
- \frac{1}{2} W_{nl}^{i} C_{np}^{-1} W_{pj}^{i} m_{l}^{i,t} \\
&+ \frac{1}{2} W_{nj}^{i} C_{np}^{-1} W_{pq}^{r} m_{q}^{r,t}
+ \frac{1}{2} W_{nl}^{r} C_{np}^{-1} W_{pj}^{i} m_{l}^{r,t} \\
&+ \frac{1}{2} W_{nj}^{i} C_{np}^{-1} W_{pq}^{i} \delta_{jq}
- A_{jl}^{i} m_{l}^{i,t-1}
- A_{lj}^{i} m_{l}^{i,t+1}
+ \lambda_{i,t} \\
= &
\ln m_{j}^{i,t} + 1
- y_{p}^{t} C_{pn}^{-1} W_{nj}^{i}
- W_{nl}^{i} C_{np}^{-1} W_{pj}^{i} m_{l}^{i,t}
+ W_{nl}^{r} C_{np}^{-1} W_{pj}^{i} m_{l}^{r,t} \\
&+ \frac{1}{2} W_{nj}^{i} C_{np}^{-1} W_{pq}^{i} \delta_{jq}
- A_{jl}^{i} m_{l}^{i,t-1}
- A_{lj}^{i} m_{l}^{i,t+1}
+ \lambda_{i,t}
\end{align}
with a special version for the first in the time sequence:

\begin{align}
\frac{\partial\mathcal{KL}}{\partial m_{j}^{i,t=1}} = &
\ln m_{j}^{i,1} + 1
- y_{p}^{1} C_{pn}^{-1} W_{nj}^{i}
- W_{nl}^{i} C_{np}^{-1} W_{pj}^{i} m_{l}^{i,t}
+ W_{nl}^{r} C_{np}^{-1} W_{pj}^{i} m_{l}^{r,1} \\
&+ \frac{1}{2} W_{nj}^{i} C_{np}^{-1} W_{pq}^{i} \delta_{jq}
- A_{lj}^{i} m_{l}^{i,2}
- \ln \pi_{j}^{i}
+ \lambda_{i,1}
\end{align}
and the $\lambda$ terms supply the probability condition.  Defining
$\vec{\hat{y}}^{\,t} = W^{i} \cdot \vec{m}^{\,i,t}$, we arrive at
\begin{equation}
m_{j}^{i,t} = 
\sigma\left(C_{pn}^{-1} W_{nj}^{i} \left(y_{p}^{t} - \hat{y}_{p}^{t} \right)
+ W_{nl}^{i} C_{np}^{-1} W_{pj}^{i} m_{l}^{i,t}
- \frac{1}{2} W_{nj}^{i} C_{np}^{-1} W_{pq}^{i} \delta_{jq}
- 1
+ A_{jl}^{i} m_{l}^{i,t-1}
+ A_{lj}^{i} m_{l}^{i,t+1} \right)
\label{eq:m}
\end{equation}

%\begin{align}
%m_{j}^{i,t} = 
%\sigma\left(\right. & \left.\frac{1}{2} y_{p}^{t} C_{pn}^{-1} W_{nj}^{i}
%- W_{nl}^{r} C_{np}^{-1} W_{pj}^{i} m_{l}^{r,t}\right. \\
%& \left. - \frac{1}{2} W_{nj}^{i} C_{np}^{-1} W_{pq}^{i} \delta_{jq}
%- 1
%+ A_{jl}^{i} m_{l}^{i,t-1}
%+ A_{lj}^{i} m_{l}^{i,t+1} \right)
%\end{align}

\begin{equation}
m_{j}^{i,1} = 
\sigma\left(C_{pn}^{-1} W_{nj}^{i} \left( y_{p}^{1} - \hat{y}_{p}^{1}\right)
+ W_{nl}^{i} C_{np}^{-1} W_{pj}^{i} m_{l}^{i,1}
- \frac{1}{2} W_{nj}^{i} C_{np}^{-1} W_{pq}^{i} \delta_{jq}
- 1
+ A_{lj}^{i} m_{l}^{i,2}
+ \ln \pi_{j}^{i} \right)
\end{equation}
%\begin{align}
%m_{j}^{i,1} = 
%\sigma\left(\right. & \left.\frac{1}{2} y_{p}^{1} C_{pn}^{-1} W_{nj}^{i}
%- W_{nl}^{r} C_{np}^{-1} W_{pj}^{i} m_{l}^{r,1}\right. \\
%& \left.- \frac{1}{2} W_{nj}^{i} C_{np}^{-1} W_{pq}^{i} \delta_{jq}
%- 1
%+ \ln \pi_{j}^{i} \right)
%\end{align}
where $\sigma$ is the softmax function, enforcing the probability condition
as imposed by solving the $\lambda_{i}$ Lagrangian multiplier equations.
These are the set of fixed-point equations for finding the mean field
parameters $m_{j}^{i,t}$.

We implement this fixed-point maximization by updating the $k$-components,
for each randomly selected $(t, d)$ pair, all enclosed inside an iteration
loop that checks the KLD convergence as an exit criterion.  Inside the
$(t, d)$ loop we have the following:
\begin{tcolorbox}
\begin{lstlisting}
wm = np.einsum('dok,dk', W, m[t])
y_err = x[t] - wm
log_m_new = np.zeros(shape=(self.k,))
for k in range(self.k):
    am = A[d, k, :].dot(m[t-1, d, :])
    ma = m[t+1, d, :].dot(A[d, :, k])

    if t == 0:
        am = np.log(pi[d, k])
    if t == self.T-1:
        ma = 0

    log_m_new[k] = W[d, :, k].dot(C_inv.dot(y_err)) \
                 + W[d, :, k].dot(C_inv.dot(W[d].dot(m[t, d, :])))\
                 - 1/2 * W[d, :, k].dot(C_inv.dot(W[d, :, k])) \
                 - 1 + ma + am
m[t, d, :] = np.clip(softmax(log_m_new),
                     zero_probability,
                     1-zero_probability)
m[t, d, :] /= m[t, d, :].sum()
\end{lstlisting}
\end{tcolorbox}
\noindent
where we first compue the difference between the data and the estimate
of $\vec{y}$.  We then update each $k$-component of $m$, first
checking the edge cases, selecting $\pi$ for the
$A_{jl}^{i} m_{l}^{i,t-1}$ term (\lstinline|am|) in the case of $t=0$ and
setting the $A_{lj}^{i} m_{l}^{i,t+1}$ (\lstinline|ma|) to $0$ in the $t=T-1$
case, since there is no $T+1$ term in the derivation of the $m$ equation.
We then update $m$ with code that directly reflects Eq.~\ref{eq:m}. Finally,
we optionally clip the numerical values, depending on the value of
\lstinline|zero_probability|, and normalize to 1.

\section{Detailed Gibbs Sampling Expectation Estimation}

In Gibbs sampling we sample the states from the conditional probability distribution:

\begin{align}
S_{t}^{i} & \sim  P(S_{t}^{i}|\{S_{t}^{\widehat{i}}\}, S_{t-1}^{i}, S_{t+1}^{i}, Y_t)
= P(S_{t}^{i}| {\rm MB}) \\
& \propto P(S_{t}^{i} | S_{t-1}^{i}) P(S_{t+1}^{i} | S_{t}^{i}) P(Y_t | \{ S_{t} \})
\label{eq:ps_gibbs}
\end{align}
where the hatted chain is excluded from the set~\cite{Ghahramani1997},
and ${\rm MB}$ stands for the Markov blanket around $S_{t}^{i}$. To carry
out the Gibbs sampling procedure, for each $t$ we randomly draw each chain hidden
state according to its conditional distribution.  We then set those $t$ chain
states to the random draw values.  In order to calculate the conditional probability
distribution of a particular chain at a particular time, we have in inner $k$-loop
that calculates the probability for each possible $k$ value. In code, for each value
of $t$ and $d$, we update all $k$ state probability values as follows:
\begin{tcolorbox}
\begin{lstlisting}
old_s = self.s.copy()  # store prior to `d` updates
for d in range(self.d):
    s = old_s.copy()
    for k in range(self.k):
        s[t, d, :] = 0
        s[t, d, k] = 1

        # Edge case -- end of sequence
        if t == self.T-1:
            A_tp1 = 1
        else:
            state_tp1_idx = np.argmax(s[t+1, d, :])
            A_tp1 = np.exp(self.A[d, state_tp1_idx, k])

        # Edge case -- beginning of sequence
        if t == 0:
            A_tm1 = self.pi[d, k]
        else:
            state_tm1_idx = np.argmax(s[t-1, d, :])
            A_tm1 = np.exp(self.A[d, k, state_tm1_idx])

        y_mu = np.einsum('dok,dk', self.W, s[t, :, :])
        pyt = scs.multivariate_normal.pdf(x[t, :], y_mu, self.C)

        self.ps[i, t, d, k] = A_tm1 * pyt * A_tp1 \
                              + self.zero_probability

    # Randomly draw from the conditional distribution
    idx = np.random.choice(range(self.k), p=self.ps[i, t, d, :])

    # Update to drawn state
    self.s[t, d, :] = 0
    self.s[t, d, idx] = 1
\end{lstlisting}
\end{tcolorbox}
\noindent
Since we are trying to calculate the conditional probability,
Eq.~\ref{eq:ps_gibbs}, for all values of $k$ for each $t$ and $d$,
we first check the edge cases. If we are at
the end of the sequence $t=T-1$, then we remove the probability
$P(S_{T+1}^{i}|S_{t}^{i})$, setting \lstinline|A_tp1| to 1.
If we are at the beginning of the sequence we set $P(S_{1}^{i}|S_{0}^{i})$
to $\pi$. Otherwise, we set \lstinline|A_tm1| and \lstinline|A_tp1| to their
appropriate exponentiated $A$ values.  We then calculate $P(Y_t|S_{t})$
using the current hidden states, labeled \lstinline|pyt|, and fill out
the probability matrix \lstinline|ps|, using Eq.~\ref{eq:ps_gibbs}.
We store these hidden state trajectories and probability trajectories
over all iterations in \lstinline|states[i, t, d, k]| and
\lstinline|ps[i, t, d, k]|, respectively.

We then proceed to calculate the hidden state expectation values, based on the
traces of states and probabilities.  We estimate $\langle S_{t}^{i} \rangle$
via averaging the state values over iterations:

\begin{equation}
\langle S_{t}^{i} \rangle = \frac{1}{N_G} \sum_{n = 1}^{N_G}
S_{t}^{i, (n)}
%\; P(S_{t}^{i, (n)}| {\rm MB})
%\langle S_{t}^{i} \rangle = \frac{1}{N_G} \sum_{n = 1}^{N_G}
%S_{t}^{i, (n)} P(S_{t}^{i, (n)}|\{S_{t}^{\widehat{i}, (n)}\}, S_{t-1}^{i, (n)}, S_{t+1}^{i, (n)}, Y_t)
\end{equation} 
where $N_G$ is the total number of Gibbs iterations, and the index in parentheses, $(n)$,
refers to the $n$th iteration value, effecting an average over iterations.
The code implementation is very short:
\begin{tcolorbox}
\begin{lstlisting}
for i in range(len(self.states)):
    s_exp += self.states[i]
s_exp /= len(self.states)
\end{lstlisting}
\end{tcolorbox}
\noindent
To calculate the same-time-different-chain state expectation, we average using
both conditional probabilities:

\begin{equation}
\langle S_{t}^{i} S_{t}^{j} \rangle = \frac{1}{N_G} \sum_{n = 1}^{N_G}
S_{t}^{i, (n)}
\; S_{t}^{j, (n)}
%\; P(S_{t}^{i, (n)}| {\rm MB})
%\; P(S_{t}^{j, (n)}| {\rm MB})
%S_{t}^{i, (n)} P(S_{t}^{i, (n)}|\{S_{t}^{\widehat{i}, (n)}\}, S_{t-1}^{i, (n)}, S_{t+1}^{i, (n)}, Y_t)
%S_{t}^{j, (n)} P(S_{t}^{j, (n)}|\{S_{t}^{\widehat{j}, (n)}\}, S_{t-1}^{j, (n)}, S_{t+1}^{j, (n)}, Y_t)
\end{equation} 
The code implementation does this via $k$-space outer product, averaged over
iterations:

\begin{tcolorbox}
\begin{lstlisting}
for i in range(len(self.states)):
    for t, d1 in td:
        for d2 in range(self.d):
            ss_exp[t, d1, d2] += np.outer(self.states[i, t, d1, :],
                                          self.states[i, t, d2, :])
ss_exp /= len(self.states)
\end{lstlisting}
\end{tcolorbox}
\noindent
Lastly, to calculate the same-chain-different-time state expectation,
we average using the conditional probabilities and the transition
probability:
\begin{equation}
\langle S_{t}^{i} S_{t-1}^{i} \rangle = \frac{1}{N_G} \sum_{n = 1}^{N_G}
S_{t}^{i, (n)}
\; S_{t-1}^{i, (n)}
%\; P(S_{t}^{i, (n)}| {\rm MB})
%\; P(S_{t-1}^{i, (n)}| {\rm MB})
%\; P(S_{t}^{i, (n)}| S_{t-1}^{i, (n)})
%S_{t}^{i, (n)} P(S_{t}^{i, (n)}|\{S_{t}^{\widehat{i}, (n)}\}, S_{t-1}^{i, (n)}, S_{t+1}^{i, (n)}, Y_t)
%S_{t-1}^{i, (n)} P(S_{t-1}^{i, (n)}|\{S_{t-1}^{\widehat{i}, (n)}\}, S_{t-2}^{i, (n)}, S_{t}^{i, (n)}, Y_{t-1})
\end{equation} 
In code, this is a similar element-wise multiplied outer product,
taking care to skip the first
\lstinline|t=0| entry, then averaging over iterations:

\begin{tcolorbox}
\begin{lstlisting}
for i in range(len(self.states)):
    for t, d in td:
        if t == 0:
            continue
        sstm1_exp[t, d] += np.outer(self.states[i, t, d, :],
                                    self.states[i, t-1, d, :])
sstm1_exp /= len(self.states)
\end{lstlisting}
\end{tcolorbox}
\noindent

\section{Detailed Structured Variational Approximation Estimation}

Since the derivation is similar to the Mean Field case, we review the Structured Variational Approximation
(SVA) estimation \cite{Ghahramani1997} in our notation, leaving out some details.  Similar to the Mean
Field derivation we use the probability distribution:

\begin{equation}
\widetilde{P} = \frac{1}{\mathcal{Z}_{\rm SVA}} \prod_{d=1}^{D} \widetilde{P}(S_{1}^{d}|\theta)
                \prod_{t=2}^{T} \widetilde{P}(S_{t}^{d} | S_{t-1}^{d} \theta)
\end{equation}

\begin{equation}
\widetilde{P} = \frac{1}{\mathcal{Z}_{\rm SVA}} \prod_{d=1}^{D} \prod_{k}^{K} (h_{1k}^{d} \pi_{k}^{d})^{S_{1k}^{d}}
                \prod_{t=2}^{T} \prod_{i}^{K} \left(h_{ti}^{d} \prod_{j}^{K} (P_{ij}^{d})^{S_{t-1,j}^{d}}\right)^{S_{ti}^{d}}
\end{equation}
The only unspecified piece here is the normalization $\mathcal{Z}_{\rm SVA}$.  This is determined as follows:
\begin{equation}
\mathcal{Z}_{\rm SVA} = \sum_{\{S\}}\prod_{d=1}^{D} \prod_{k}^{K} (h_{1k}^{d} \pi_{k}^{d})^{S_{1k}^{d}}
                \prod_{t=2}^{T} \prod_{i}^{K} \left(h_{ti}^{d} \prod_{j}^{K} (P_{ij}^{d})^{S_{t-1,j}^{d}}\right)^{S_{ti}^{d}}
\end{equation}
We start by looking at the first product on $K$. This is summed over all values of $S_{1k}^{d}$, but for each $d$,
only one of the $k$ values results in $S_{1k}^{d} = 1$, this means we can substitute the $S_{1k}^{d}$ part of the
outermost sum with

\begin{equation}
\sum_{\{S\}} \prod_{k}^{K} (h_{1k}^{d} \pi_{k}^{d})^{S_{1k}^{d}} \rightarrow \sum_{k}^{K} h_{1k}^{d} \pi_{k}^{d}
\end{equation}
Similarly, for the $\prod_{i}^{K}$ and $\prod_{j}^{K}$ these terms are not $1$ for specific selections of $i$, and $j$,
where $S_{ti}^{d} = 1$ and $S_{t-1,j}^{d}$, but since we're summing over all realizations, we can substitute

\begin{equation}
\sum_{\{S\}} \prod_{i}^{K} \left(h_{ti}^{d} \prod_{j}^{K} (P_{ij}^{d})^{S_{t-1,j}^{d}}\right)^{S_{ti}^{d}} \quad
  \rightarrow \quad \sum_{i}^{K} h_{ti}^{d} \sum_{j}^{K} P_{ij}^{d} = 1 \quad ,
\end{equation}
where the last equality results from the normalization of both $P$ and $h$.  So our final expression for
$\mathcal{Z}_{\rm SVA}$ is

\begin{equation}
\mathcal{Z}_{\rm SVA} = \prod_{d=1}^{D} \sum_{k}^{K} h_{1k}^{d} \pi_{k}^{d} \quad .
\end{equation}
Taking the logarithm of our SVA distribution $\widetilde{P}$ and plugging into the equation for the KLD, we get

\begin{align}
\mathcal{KL} = &
\sum_{t,i,j}^{T,d,k} \langle s_{j}^{i,t} \rangle_{\widetilde{P}} \ln h_{j}^{i,t}
- \ln \mathcal{Z}_{\rm SVA}
+ \frac{1}{2} \sum_{t=1}^{T} y_{i}^{t} C_{ij}^{-1} y_{j}^{t}
- \sum_{t=1}^{T} W_{nl}^{i} \langle s_{l}^{i,t}\rangle_{\widetilde{P}} C_{np}^{-1} y_{p}^{t} \\
&+ \frac{1}{2} \sum_{t=1}^{T} W_{nl}^{i} C_{np}^{-1} W_{pq}^{j}
  \langle s_{l}^{i,t} s_{q}^{j,t} \rangle_{\widetilde{P}}
+ \ln \mathcal{Z} \\
\end{align}
Taking the derivative with respect to $\ln h$, we find the update equation for $h$ (Appendix D of~\cite{Ghahramani1997}):

\begin{equation}
h_{k}^{d,t} = \exp\left[W_{nk}^{d} C_{np}^{-1} \left(Y_{\rm err}^{(d)}\right)_{p}^{t} - \frac{1}{2} W_{nk}^{d} C_{np}^{-1} W_{pk}^{d}\right]
\end{equation}
where
\begin{equation}
\left(Y_{\rm err}^{(d)}\right)_{o}^{t} = Y_{o}^{t} - \sum_{m \neq d}^{D} W_{ok}^{m} \langle s_{k}^{m,t} \rangle \quad \quad .
\end{equation}

Each iteration of the SVA routine is implemented as follows:
\begin{tcolorbox}
\begin{lstlisting}
for t, d in sorted(td, key=lambda x: np.random.random()):
    y_err = np.zeros(shape=(self.o,))
    ws = 0
    for dm in range(self.d):
        if dm == d:
            continue
        ws += W[dm].dot(s_exp[t, dm])
    y_err = x[t] - ws

    # Update and normalize the vector h[t, d, :]
    log_h_new = np.einsum('ok,o->k', W[d], C_inv.dot(y_err)) \
                - 1/2 * np.einsum('ok,op,pk->k', W[d], C_inv, W[d])
    h[t, d, :] = np.clip(softmax(log_h_new), zero_probability, 1-zero_probability)
    h[t, d, :] /= h[t, d, :].sum()

s_exp, ss_exp, sstm1_exp = self.forward_backward(h)
\end{lstlisting}
\end{tcolorbox}
\noindent
where we update and normalize the \lstinline|h[t, d, :]| probability vector for each value of
\lstinline|t, d|, as per the above equations. Finally we update the expectation values based on
the new \lstinline|h|, using the regular forward-backward algorithm on each chain separately.

\section{FHMM Viterbi Algorithm}
\label{appendix:fhmm_viterbi_algorithm}

We extend the Viterbi algorithm for HMMs in the natural way to FHMMs, following
the logic of~\cite{Nefian2002}, linking the notation of the algorithm for HMMs
from Rabiner~\cite{Rabiner1989}, to our notation. The notational link is
provided by

\begin{equation}
b_{j}(O_{t}) = P(Y_{t}|S_{t}=s_{t}^{(j)})
\end{equation}

and

\begin{equation}
a_{ij} = P(q_t=j|q_{t-1}=i) \rightarrow P(S_{t}=s_{t}^{(j)}|S_{t-1}=s_{t-1}^{(i)})
\end{equation}
Note the $a_{ij}$ indices are flipped from our convention. Also, importantly,
`$j$' and `$i$' refer to a particular hidden state assignment, so for us,
since $S_{t}$ is representing
$d$ chains and $k$ states, this is a $d \times k$ set of values. Instead
of using $j$ directly as on the left-hand side we use $s^{(j)}$ to specify
the realization, for clarity.  Due to the graph dependency the probability
decomposes as

\begin{equation}
P(S_{t}=s_t^{(j)}|S_{t-1}=s_{t-1}^{(i)}) = \prod_{l=1}^{d} P(s_{t}^{l, (j)}|s_{t-1}^{l, (i)})
\end{equation}
with a similar expansion relevant for $\pi$.  With this mapping in hand the recursion relation of \cite{Rabiner1989} becomes:

\begin{equation}
\delta_{t}(i) = \left[\max_j \delta_{t-1}(j) \prod_{l=1}^{d} P(s_{t}^{l, (i)}|s_{t-1}^{l, (j)})\right] P(Y_{t}|S_{t}=s_{t}^{(i)}; \phi)
\end{equation}

We now write out the algorithm of \cite{Rabiner1989} using our notation, with
the minor alteration and additional complexity multiplying probabilities
across chains.

1) Initialization

\begin{equation}
\delta_1(i) = P(S_{1}=s_{1}^{(i)}) P(Y_{1}|S_{1}=s_{1}^{(i)}), \quad \forall i \in {\rm realizations} \,\,.
\end{equation}
\begin{equation}
\psi_{1}(i) = 0\,\, .
\end{equation}

\begin{tcolorbox}
\begin{lstlisting}
# Initialize delta (just like alpha)
delta = np.zeros(shape=(self.T, self.k**self.d))
psi = np.zeros(shape=(self.T, self.k**self.d), dtype=np.int)
for i in range(realizations.shape[1]):
    pi = 1
    for d in range(self.d):
        pi *= self.pi[d, realizations[d, i]]

    delta[0, i] = pi * py[0, i] + eps
    psi[0, i] = 0
\end{lstlisting}
\end{tcolorbox}
\noindent

2) Recursion

\begin{equation}
\delta_{t}(i) = \max_j \left[\delta_{t-1}(j) \prod_{l=1}^{d} P(s_{t}^{l, (i)}|s_{t-1}^{l, (j)})\right] P(Y_{t}|S_{t}=s_{t}^{(i)}; \phi)
\end{equation}

\begin{equation}
\psi_{t}(i) = \argmax_j \left[\delta_{t-1}(j) \prod_{l=1}^{d} P(s_{t}^{l, (i)}|s_{t-1}^{l, (j)})\right]
\end{equation}

\begin{tcolorbox}
\begin{lstlisting}
for t in range(1, self.T):
    for j in range(realizations.shape[1]):
        prob_j = 1
        for d in range(self.d):
            prob_j *= np.exp(self.A)[d, realizations[d, j], realizations[d, :]]
        delta[t, j] = np.max(delta[t-1] * prob_j) * py[t, j] + eps
        psi[t, j] = np.argmax(delta[t-1] * prob_j)

    delta[t, :] /= delta[t, :].sum()
\end{lstlisting}
\end{tcolorbox}
\noindent
In the above code you can see the additional inner-most for loop, necessary
since this is a FHMM with many hidden chains.  When $d=1$, the for loop would
disappear and the computation would reduce to the HMM version.

3) Termination

\begin{equation}
P^{*} = \max_i \delta_{T}(i)
\end{equation}

\begin{equation}
q_{T}^{*} = \argmax_i \delta_{T}(i)
\end{equation}
Which is trivially represented in code:

\begin{tcolorbox}
\begin{lstlisting}
p_star = np.max(delta[self.T-1])
q_star = np.zeros(shape=self.T, dtype=np.int)
q_star[self.T-1] = int(np.argmax(delta[self.T-1]))
\end{lstlisting}
\end{tcolorbox}
\noindent

4) Path backtracking

\begin{equation}
q_{t}^{*} = \psi_{t+1}(q_{t+1}^{*})
\end{equation}
Since $\psi$ is holding the previous most likely realization given a current
realization, we start with the $q_{T}^{*}$ realization and work backwards.
This is a straightforward transcription in code:

\begin{tcolorbox}
\begin{lstlisting}
for t in reversed(range(self.T-1)):
    q_star[t] = int(psi[t+1, q_star[t+1]])
\end{lstlisting}
\end{tcolorbox}
\noindent
Finally, we take this trajectory through realization space and turn that into
a matrix of occupied states:

\begin{tcolorbox}
\begin{lstlisting}
states = np.zeros(shape=(self.T, self.d, self.k))

for t in range(self.T):
    for d in range(self.d):
        states[t, d, realizations[d, q_star[t]]] = 1
\end{lstlisting}
\end{tcolorbox}
\noindent

\section{Detailed Hessian computation}
\label{appendix:detailed_hessian}

Here we outline our calculation of the Hessian of the log likelihood for our FHMM,
analogous to Appendix A of~\cite{Aittokallio1999}.  We also substitute all model
constraints, such that we vary only with respect to the independent parameters.

\subsection{Preliminary}
The likelihood is equal to
\begin{equation}
    P(\{Y\}| \phi) = \sum_{\{S_{T}\}} P(S_{T}^{1}, ..., S_{T}^{d}, Y_{1}, ..., Y_{T} | \phi) \
     = \sum_{\{S_{T}\}} \alpha_{T}
\end{equation}
where $\{S_{T}\}$ in the sum indicates a sum over all $d$ hidden state configurations at (final) time $T$.
We can extract the likelihood from the forward recurrence relation:
\begin{equation}
    \alpha_{t} = P(Y_t | \{S_t\}) \prod_{i=1}^{d} \sum_{\{S_{t-1}\}} P(S_{t}^{i}|S_{t-1}^{i}) \alpha_{t-1}
\end{equation}
First, we normalize $\alpha$'s in the recurrence relation such that our recurrence relation looks like the
following
(we'll use $\widetilde{\alpha}$ to indicate not yet divided by $c$)
\begin{equation}
    \widetilde{\alpha}_{t} = P(Y_t | \{S_t\}) \prod_{i=1}^{d} \sum_{\{S_{t-1}\}} P(S_{t}^{i}|S_{t-1}^{i}) \widehat{\alpha}_{t-1}
\label{eq:alpha_recursion}
\end{equation}

\begin{equation}
    \widehat{\alpha}_{t-1} = \widetilde{\alpha}_{t-1} / c_{t-1}
\end{equation}
where $c_{t-1} = \sum_{\{S_{t-1}\}} \widetilde{\alpha}_{t-1}$.
In the above, $\widetilde{\alpha}_{t}$ can be thought of as a function of possible $S_{t}^{i}$ (binary) values.
Or, when programming, a vector of length $d^{k}$ with entries containing an evaluation of
$\widetilde{\alpha}_{t}$ for each configuration of $S_{t}$.
Calculating the forward relation with this normalization makes the numerical routine more stable and
also allows for an easy method to track the $c$'s and calculate the log likelihood.
\begin{align}
    \widetilde{\alpha}_{T} &= P(Y_T | \{S_T\}) \prod_{i=1}^{d} \sum_{\{S_{T-1}\}} P(S_{T}^{i}|S_{T-1}^{i}) \widehat{\alpha}_{T-1} \\
    &= \left(\prod_{j=1}^{T-1} \frac{1}{c_{j}}\right)P(Y_T | \{S_T\}) \prod_{i=1}^{d} \sum_{\{S_{T-1}\}} P(S_{T}^{i}|S_{T-1}^{i}) \alpha_{T-1}
\end{align}
Now when we sum over all hidden states we get
\begin{equation}
    c_{T} = \left(\prod_{j=1}^{T-1} \frac{1}{c_{j}}\right) \sum_{\{S_{T}\}} \alpha_{T} \
    \quad \rightarrow \quad \prod_{j=1}^{T} c_{j} = P(\{Y\}| \phi) \;.
\end{equation}
This yields our final relation for the log likelihood:
\begin{equation}
    \ln \mathcal{L} = \ln P(\{Y\}| \phi) = \sum_{j=1}^{T} \ln c_{j} \;.
\label{eq:c_ll}
\end{equation}

\subsection{Exact computation}
We are interested in calculating the Hessian of the log likelihood given in Equation~\ref{eq:c_ll}:
\begin{equation}
    \frac{\partial^{2}}{\partial Y \partial X} \sum \ln c_j = \sum\left(-\frac{1}{c_j^2} \frac{\partial c_j}{\partial Y} \frac{\partial c_j}{\partial X} + \frac{1}{c_j}\frac{\partial^{2} c_j}{\partial Y \partial X} \right) \;,
\label{eq:llhess}
\end{equation}
which implies that we need to keep track of $c$ and its derivatives during the recursion.
From the recursion for $\alpha$ in Equation~\ref{eq:alpha_recursion} we can derive:
\begin{align}
\frac{\partial \widetilde{\alpha}_{t}}{\partial X} = & \
    \frac{\partial P(Y_t | \{S_t\})}{\partial X} \prod_{i=1}^{d} \sum_{\{S_{t-1}\}} P(S_{t}^{i}|S_{t-1}^{i}) \widehat{\alpha}_{t-1} \\
    & + P(Y_t | \{S_t\}) \prod_{i=1}^{d} \sum_{\{S_{t-1}\}} \frac{\partial P(S_{t}^{i}|S_{t-1}^{i})}{\partial X} \widehat{\alpha}_{t-1} \\
    & + P(Y_t | \{S_t\}) \prod_{i=1}^{d} \sum_{\{S_{t-1}\}} P(S_{t}^{i}|S_{t-1}^{i}) \left(\widehat{\frac{\partial \alpha_{t-1}}{\partial X}} - \frac{1}{c_{t-1}} \frac{\partial c_{t-1}}{\partial X} \widehat{\alpha}_{t-1}\right)
\end{align}
where we have used a hat to indicate the partial derivative normalized by $c$: $\widehat{\frac{\partial \alpha_{t-1}}{\partial X}} = \frac{1}{c_{t-1}}\frac{\partial \widetilde{\alpha}_{t-1}}{\partial X}$.  We also have
\begin{align}
    \frac{\partial^{2} \widetilde{\alpha}_{t}}{\partial Y \partial X} = & \
    \frac{\partial^{2} P(Y_t | \{S_t\})}{\partial Y \partial X} \prod_{i=1}^{d} \sum_{\{S_{t-1}\}} P(S_{t}^{i}|S_{t-1}^{i}) \widehat{\alpha}_{t-1} \\
    & + \frac{\partial P(Y_t | \{S_t\})}{\partial X} \prod_{i=1}^{d} \sum_{\{S_{t-1}\}} \frac{\partial P(S_{t}^{i}|S_{t-1}^{i})}{\partial Y} \widehat{\alpha}_{t-1} \\
    & + \frac{\partial P(Y_t | \{S_t\})}{\partial X} \prod_{i=1}^{d} \sum_{\{S_{t-1}\}} P(S_{t}^{i}|S_{t-1}^{i}) \left(\widehat{\frac{\partial \alpha_{t-1}}{\partial Y}} - \frac{1}{c_{t-1}} \frac{\partial c_{t-1}}{\partial Y} \widehat{\alpha}_{t-1}\right) \\
    & + \frac{\partial P(Y_t | \{S_t\})}{\partial Y} \prod_{i=1}^{d} \sum_{\{S_{t-1}\}} \frac{\partial P(S_{t}^{i}|S_{t-1}^{i})}{\partial X} \widehat{\alpha}_{t-1} \\
    & + P(Y_t | \{S_t\}) \prod_{i=1}^{d} \sum_{\{S_{t-1}\}} \frac{\partial^{2} P(S_{t}^{i}|S_{t-1}^{i})}{\partial Y \partial X} \widehat{\alpha}_{t-1} \\
    & + P(Y_t | \{S_t\}) \prod_{i=1}^{d} \sum_{\{S_{t-1}\}} \frac{\partial P(S_{t}^{i}|S_{t-1}^{i})}{\partial X} \left(\widehat{\frac{\partial \alpha_{t-1}}{\partial Y}} - \frac{1}{c_{t-1}} \frac{\partial c_{t-1}}{\partial Y} \widehat{\alpha}_{t-1}\right) \\
    & + \frac{\partial P(Y_t | \{S_t\})}{\partial Y} \prod_{i=1}^{d} \sum_{\{S_{t-1}\}} P(S_{t}^{i}|S_{t-1}^{i})\
        \left(\widehat{\frac{\partial \alpha_{t-1}}{\partial X}} \
              - \frac{1}{c_{t-1}} \frac{\partial c_{t-1}}{\partial X} \widehat{\alpha}_{t-1}\right) \\
    & + P(Y_t | \{S_t\}) \prod_{i=1}^{d} \sum_{\{S_{t-1}\}} \frac{\partial P(S_{t}^{i}|S_{t-1}^{i})}{\partial Y} \
        \left(\widehat{\frac{\partial \alpha_{t-1}}{\partial X}} \
              - \frac{1}{c_{t-1}} \frac{\partial c_{t-1}}{\partial X} \widehat{\alpha}_{t-1}\right) \\
    & + P(Y_t | \{S_t\}) \prod_{i=1}^{d} \sum_{\{S_{t-1}\}} P(S_{t}^{i}|S_{t-1}^{i}) \\
    & \times \left(\widehat{\frac{\partial^{2} \alpha_{t-1}}{\partial Y \partial X}} - \frac{1}{c_{t-1}} \frac{\partial c_{t-1}}{\partial Y} \widehat{\frac{\partial \alpha_{t-1}}{\partial X}} + \frac{1}{c_{t-1}^{2}} \frac{\partial c_{t-1}}{\partial Y} \frac{\partial c_{t-1}}{\partial X} \widehat{\alpha}_{t-1} - \frac{1}{c_{t-1}} \frac{\partial^{2} c_{t-1}}{\partial Y \partial X} \widehat{\alpha}_{t-1} \right. \\
    & \quad - \left.\frac{1}{c_{t-1}} \frac{\partial c_{t-1}}{\partial X} \left(\widehat{\frac{\partial \alpha_{t-1}}{\partial Y}} - \frac{1}{c_{t-1}} \frac{\partial c_{t-1}}{\partial Y} \widehat{\alpha}_{t-1}\right)\right)
\end{align}
where we have similarly defined $\widehat{\frac{\partial^{2} \alpha_{t-1}}{\partial Y \partial X}} = \frac{1}{c_{t-1}}\frac{\partial^{2} \widetilde{\alpha}_{t-1}}{\partial Y \partial X}$.
From the above two equations we have
\begin{equation}
    \frac{\partial c_t}{\partial X} = \sum_{\{S_t\}} \frac{\partial \widetilde{\alpha}_{t}}{\partial X} \
    \quad {\rm and} \quad \frac{\partial^{2} c_t}{\partial Y \partial X} = \sum_{\{S_t\}} \frac{\partial^{2} \widetilde{\alpha}_{t}}{\partial Y \partial X}
\end{equation}
Now, just as we tracked $c$ in order to calculate the log likelihood, we additionally
track $\frac{\partial c_t}{\partial X}$ and $\frac{\partial^{2} c_t}{\partial Y \partial X}$,
in order to compute the Hessian via Equation \ref{eq:llhess}.
In many cases the recursion expressions simplify substantially. For example for the $X = W$ and $Y = W$
case, only terms 1, 3, 7, and 9 contribute, since $P(S_{t}^{i}|S_{t-1}^{i})$ has no $W$ dependence.

\subsection{Implementation}

We have the equations to calculate $\widehat{\alpha}$'s and their derivatives; and we can track
$c$'s and their derivatives. The only pieces left to show explicit calculations for are the initializations
and the remaining derivatives within the recursion: first and second derivatives of $P({Y_{t}}|\phi)$
with respect to
$W$ and $C$; first and second derivatives of $\prod P(\{S_{t}\} | \{S_{t-1}\})$ with respect to $A$;
and the derivative of the initial distribution with respect to $\pi$.

\subsubsection{Preliminary}
\label{appendix:detailed_hessian:preliminary}
We use some convenience mappings \lstinline|realizations| and \lstinline{k_contrib} to help carry out
the calculations -- they are calculated upfront and cached for repeated use.
The mapping \lstinline|realizations[idx_d, i]| is an array with the first index indicating the hidden
chain, and the second index indicating the configuration of hidden states.
For example, \lstinline|realizations[1, 2]| having an entry value of 3 means that in chain 1
in configuration (or realization) 2 is in state index 3 ($S_{t, k}^{1} = [0, 0, 0, 1]$).  (Python indexes
starting from zero.) Specifically, for $d = 2$ and $k = 2$

\[
{\rm realizations}[:, :] = 
\begin{bmatrix}
    0 & 0 & 1 & 1 \\
    0 & 1 & 0 & 1
\end{bmatrix} \; .
\]
This implies that we can set the hidden state values at time \lstinline|t| to a specific
realization \lstinline|r| via

\begin{tcolorbox}
\begin{lstlisting}
s_t = np.zeros(shape=(D, K))
for idx_d in range(D):
    s_t[idx_d, realizations[idx_d, r]] = 1
\end{lstlisting}
\end{tcolorbox}
\noindent
This construction of a realization's hidden state representation is used quite often.

The mapping \lstinline|k_contrib[idx_d, k]| is an array (or sometimes represented as a dictionary)
which takes the chain in the first position and the state index in the second position.
The value at this location is a list of realization indices having that state index in that chain.
For example, in the $d = 2$ and $k = 2$ example above, \lstinline|k_contrib[1, 1] = [1, 3]| (second chain and
second state index occur in the second and fourth realization).
Additionally, this implies that if we need a mask for all realizations having chain \lstinline|d| with state
index \lstinline|l| set (an array with the same length as the number of realizations, with 1's in the
positions that match the criteria and 0's elsewhere) we can use the following 

\begin{tcolorbox}
\begin{lstlisting}
l_indices_contrib = np.zeros(shape=realizations.shape[1])
l_indices_contrib[k_contrib[d, l]] = 1
\end{lstlisting}
\end{tcolorbox}

In addition to $\alpha$, we will also store values of $P(Y_{t}|S_{t}^{1}, ..., S_{t}^{d}; \phi)$
(denoted by \lstinline|py|) in an array of shape $(T, D^{K})$.  This is computed for each \lstinline|t|
and realization \lstinline|r|, filled out as follows (\lstinline|x[t, :]| being the sample at
time \lstinline|t|):
\begin{tcolorbox}
\begin{lstlisting}
y_mu = np.einsum('dok,dk', W, s_t[:, :])
py[t, i] = scs.multivariate_normal.pdf(x[t, :], y_mu, C)
\end{lstlisting}
\end{tcolorbox}
\noindent
Values of $\prod P(S_{t}|S_{t-1})$, denoted by \lstinline|prob_r|, will be computed for each \lstinline|t| and
realization \lstinline|r| corresponding to a specific configuration of $S_{t}$, and will have shape
$D^{K}$ corresponding to the possible configurations of $S_{t-1}$, to be summed:

\begin{tcolorbox}
\begin{lstlisting}
prob_r = np.ones(shape=realizations.shape[1])
for idx_d in range(D):
    temp = np.exp(A)[idx_d, realizations[idx_d, r], :]
    prob_r *= temp[realizations[idx_d, :]]
\end{lstlisting}
\end{tcolorbox}

\subsubsection{Initialization}
We initialize the recursion according to
\begin{equation}
    \alpha_{1} = P(Y_{1}|S_{1}^{1}, ..., S_{1}^{d}; \phi) \prod_{i=1}^{d} P(S_{1}^{i}) 
\end{equation}
In code this is the following (for each realization \lstinline|r|)
\begin{tcolorbox}
\begin{lstlisting}
joint_pi = 1
for idx_d in range(D):
    joint_pi *= pi[idx_d, realizations[idx_d, r]]
alpha[0, r] = joint_pi * py[0, r] + eps
\end{lstlisting}
\end{tcolorbox}
\noindent
Derivatives are similarly initialized -- for example, for $\frac{\partial^{2} \alpha}{\partial W \partial \pi}$:
\begin{tcolorbox}
\begin{lstlisting}
d2alphadwdpi[0, r] = djoint_dpi * dpydw[0, r]
\end{lstlisting}
\end{tcolorbox}
\noindent
where the derivatives are calculated as shown below.

\subsubsection{$W$ canonical form and constraint}
There is an ambiguity to the specification of $W$.  For simplicity we ignore the $o$ index
-- this transformation works for each $o$ value.  The inner product with $s$ yields
\begin{equation}
    \sum_{i=1}^{d} W^{i} \cdot \vec{s}_{t}^{\, i} =
      (W_{k}^{1}  + \mu) s_{kt}^{\, 1} + \cdots + (W^{i} - \mu_{k}) s_{kt}^{\, i} + \cdots + (W_{k}^{d} - \mu_{d}) s_{kt}^{\, d}
\end{equation}
where $\mu = \sum \mu_{i}$ and $\mu_i = \sum_{j=1}^{k} W_{j}^{i} / k$, the mean on the
$k$-axis. This yields $d-1$ constraints from the zero mean terms, which we choose to be
enacted on the $k$th element, such that $W_{k}^{i} = -\sum_{j=1}^{k-1} W_{j}^{i}$, for
each $i$ from $2$ to $d$.
This allows us to define the {\it canonically transformed} $W$, with the means added to
the first component and the other components set to zero mean.

\subsubsection{$W$ and $C$ derivatives}
The only $W$ dependence comes from the likelihood
\begin{equation}
    P(Y|\{S\},\phi) = \sqrt{\frac{(2\pi)^{-o}}{\det C}} e^{-\frac{1}{2} (\vec{y}_{t} - W\cdot\vec{s}_{t}) C^{-1} (\vec{y}_{t} - W\cdot\vec{s}_{t})^{T}}
\end{equation}

\begin{equation}
\frac{\partial P(Y_t|\{S\},\phi)}{\partial W_{ok}^{d}} = \
    \mathcal{N} \left[S_{k}^{d} C_{oa}^{-1} (\vec{y}_{t} - W \cdot \vec{S}_{t})_{a}\right] \
    = \mathcal{N} \left[S_{k}^{d} C_{oa}^{-1} (\vec{y}_{t}^{\;\rm err})_{a}\right]
\end{equation}
This is represented in the following code, looping over each \lstinline|t| and \lstinline|r|:
\begin{tcolorbox}
\begin{lstlisting}
d_constraint = s_t[d, -1] if d != 0 else 0
y_err = x[t, :] - np.einsum('dok,dk', W, s_t)
sCyWs = (s_t[d, k] - d_constraint) * C_inv[o, :].dot(y_err)
dpydw[t, r] = py[t, r] * sCyWs
\end{lstlisting}
\end{tcolorbox}
\noindent
where the constraint is enacted for \lstinline|d > 0| and on the \lstinline|k-1| element,
using \lstinline|d_constraint|.

\begin{equation}
\frac{\partial P(Y_t|\{S\},\phi)}{\partial W_{pl}^{e} \partial W_{ok}^{d}} = \
    \mathcal{N} \left[S_{l}^{e} C_{pa}^{-1} (\vec{y}_{t}^{\;\rm err})_{a} \
                      S_{k}^{d} C_{ob}^{-1} (\vec{y}_{t}^{\;\rm err})_{b} \
                      - S_{l}^{e} C_{po}^{-1} S_{k}^{d} \right]
\end{equation}
This is represented in the following code, looping over each \lstinline|t| and \lstinline|r|:
\begin{tcolorbox}
\begin{lstlisting}
d_constraint = s_t[d, -1] if d != 0 else 0
e_constraint = s_t[e, -1] if e != 0 else 0
y_err = x[t, :] - np.einsum('dok,dk', W, s_t)
sCyWs1 = (s_t[e, l] - e_constraint) * C_inv[p, :].dot(y_err)
sCyWs2 = (s_t[d, k] - d_constraint) * C_inv[o, :].dot(y_err)
sCs = (s_t[e, l] - e_constraint) * C_inv[p, o] * (s_t[d, k] - d_constraint)
d2pydwdw[t, r] = py[t, r] * (sCyWs1 * sCyWs2 - sCs)
\end{lstlisting}
\end{tcolorbox}
\noindent
where the constraint is enacted for \lstinline|d > 0| and on the \lstinline|K-1| elements,
via simple substitution using \lstinline|d_constraint| and \lstinline|e_constraint|.

Moving on, we will now calculate the $C$ derivatives. Using the relation
$\frac{\partial \det C}{\partial C} = C^{-1} \det C$
\begin{equation}
    \frac{\partial P}{\partial C_{ij}} = -\frac{1}{2} C^{-1}_{ij} \mathcal{N} + \left(-\frac{1}{2} (\vec{y}_{t} - W\cdot\vec{s}_{t})_{a} \frac{\partial C^{-1}_{ab}}{\partial C_{ij}} (\vec{y}_{t} - W\cdot\vec{s}_{t})^{T}_{b}\right) \mathcal{N}
\end{equation}
where we can use the help of the following relations:

\begin{equation}
    \frac{\partial C_{ab}^{-1}}{\partial C_{ij}} = -C_{ai}^{-1} C_{bj}^{-1} \quad {\rm and} \quad \
    \frac{\partial^{2} C_{ab}^{-1}}{\partial C_{lm} \partial C_{ij}} = \
      C_{al}^{-1} C_{im}^{-1} C_{bj}^{-1} + C_{ai}^{-1} C_{bl}^{-1} C_{jm}^{-1}
\end{equation}
to obtain

\begin{equation}
    \frac{\partial P}{\partial C_{ij}} = \frac{1}{2} \mathcal{N} \left( (\vec{y}_{t}^{\;\rm err})_{a} C_{ai}^{-1} C_{bj}^{-1} (\vec{y}_{t}^{\;\rm err})^{T}_{b} -  C^{-1}_{ij}\right)
\end{equation}
In code, this derivative is calculated as follows, for each \lstinline|t| and particular
realization \lstinline|r| :

\begin{tcolorbox}
\begin{lstlisting}
y_err = x[t, :] - np.einsum('dok,dk', W, s_t)
yCCy = y_err.dot(C_inv[:, i]) * C_inv[:, j].dot(y_err) - C_inv[i, j]
dpydc[t, r] = 1/2 * py[t, r] * yCCy
\end{lstlisting}
\end{tcolorbox}
For the second derivative we have

\begin{align}
    \frac{\partial^{2} P}{\partial C_{lm} \partial C_{ij}} = &\frac{1}{4} \mathcal{N}\
    \left((\vec{y}_{t}^{\;\rm err})_{c} C_{cl}^{-1} C_{dm}^{-1} (\vec{y}_{t}^{\;\rm err})^{T}_{d} - \
          C^{-1}_{lm}\right) \
    \left((\vec{y}_{t}^{\;\rm err})_{a} C_{ai}^{-1} C_{bj}^{-1} (\vec{y}_{t}^{\;\rm err})^{T}_{b} - \
            C^{-1}_{ij}\right) \\
    & + \frac{1}{2} \mathcal{N} \left(C_{il}^{-1} C_{jm}^{} - (\vec{y}_{t}^{\;\rm err})_{a} \
        \left[C_{al}^{-1} C_{im}^{-1} C_{bj}^{-1} + C_{ai}^{-1} C_{bl}^{-1} C_{jm}^{-1} \right] \
        (\vec{y}_{t}^{\;\rm err})_{a}\right)
\end{align}
In code, this double derivative is calculated as follows, for each \lstinline|t| and particular
realization \lstinline|r|:

\begin{tcolorbox}
\begin{lstlisting}
yCCy1 = y_err.dot(C_inv[:, l]) * C_inv[:, m].dot(y_err) - C_inv[l, m]
yCCy2 = y_err.dot(C_inv[:, i]) * C_inv[:, j].dot(y_err) - C_inv[i, j]
yCCCy1 = y_err.dot(C_inv[:, l]) * C_inv[i, m] * C_inv[:, j].dot(y_err)
yCCCy2 = y_err.dot(C_inv[:, i]) * C_inv[:, l].dot(y_err) * C_inv[j, m]

d2pydcdc[t, r] = 1/4 * py[t, r] * yCCy1 * yCCy2 \
    + 1/2 * py[t, r] * (C_inv[i, l] * C_inv[j, m] - yCCCy1 - yCCCy2)
\end{lstlisting}
\end{tcolorbox}

Finally we have the cross derivative:

\begin{align}
    \frac{\partial P(Y_t|\{S\},\phi)}{\partial W_{ok}^{d} \partial C_{}} = & \
    \frac{1}{2} \frac{\partial \mathcal{N}}{\partial W_{ok}^{d}}\
       \left((\vec{y}_{t}^{\;\rm err})_{a} C_{ai}^{-1} C_{bj}^{-1} \
             (\vec{y}_{t}^{\;\rm err})_{b} \
               -  C^{-1}_{ij}\right) \\
    & - \frac{1}{2} \mathcal{N} \
      \left[S_{k}^{d} C_{oi}^{-1} C_{bj}^{-1} (\vec{y}_{t}^{\;\rm err})_{b} \
            + (\vec{y}_{t}^{\;\rm err})_{a} C_{ai}^{-1} C_{oj}^{-1} S_{k}^{d} \right]
\end{align}
In code, this double derivative is calculated as follows, for each \lstinline|t| and particular
realization \lstinline|r|:
\begin{tcolorbox}
\begin{lstlisting}
d_constraint = s_t[d, -1] if d != 0 else 0
y_err = self.x[t, :] - np.einsum('dok,dk', self.W, s_t)
sCyWs = (s_t[d, k] - d_constraint) * self.C_inv[o, :].dot(y_err)
dpydw = self.py[t, r] * sCyWs
yCCy = y_err.dot(self.C_inv[:, i]) * self.C_inv[:, j].dot(y_err)
sCCy = (s_t[d, k] - d_constraint) * self.C_inv[o, i] \
       * self.C_inv[:, j].dot(y_err)
yCCs = y_err.dot(self.C_inv[:, i]) * self.C_inv[o, j] 
       * (s_t[d, k] - d_constraint)
d2pydwdc[t, r] = 1/2 * dpydw * (yCCy - self.C_inv[i, j]) \
                 - 1/2 * self.py[t, r] * (sCCy + yCCs)
\end{lstlisting}
\end{tcolorbox}
with the \lstinline|d_constraint| applied where appropriate.

\subsubsection{$A$ derivatives}
The $A$ derivatives are a little tricky.
Recall that the probability of transitioning from the $l$th state of $S_{t-1}^{d}$ to the
$k$th state of is $S_{t}^{d}$ is

\begin{equation}
    P((S_{t}^{d})_{k}|(S_{t-1}^{d})_{l}) = e^{A_{kl}^{d}}
\end{equation}
This implies that when taking the derivative of the product of probabilities
$\prod_{i} P(S_{t}^{i}|S_{t-1}^{i})$
for a particular realization of $S_{t-1}$ and $S_{t}$, the first derivative w.r.t
$A_{kl}^{d}$ does nothing (if the term contains the exponential of $A_{kl}^{d}$),
involves a minus sign when acting on the constraint equation (if it contains the appropriate
$A$), or
results in zero (if the term does not contain the exponential of $A_{kl}^{d}$).  Our choice
is to substitute the constraint for the $K-1$ index of $A$.

This is represented in code as follows, for each \lstinline|t| and particular
realization \lstinline|r|:
\begin{tcolorbox}
\begin{lstlisting}
dprob_rdA1 = np.ones(shape=realizations.shape[1])
k_indices = k_contrib[d, k]
d_Km1_indices = k_contrib[d, K-1]
l_indices_contrib = np.zeros(shape=self.realizations.shape[1])
l_indices_contrib[k_contrib[d, l]] = 1
for idx_d in range(D):
    temp = np.exp(A)[idx_d, realizations[idx_d, r], :]
    if idx_d == d:
        if r in k_indices:
            # Only keep terms that have d, k, l
            dprob_rdA1 *= temp[realizations[idx_d, :]] \
                          * l_indices_contrib  
        elif r in d_Km1_indices:
            temp2 = np.exp(A)[idx_d, k, :]
            dprob_rdA1 *= -temp2[realizations[idx_d, :]] \
                          * l_indices_contrib
        else:
            dprob_rdA1 *= 0
            break
    else:
        dprob_rdA1 *= temp[realizations[idx_d, :]]

\end{lstlisting}
\end{tcolorbox}
\noindent
In words: we collect the realizations that have state $k$ set and store in \lstinline|k_indices|. We also
create a mask of all realizations that contain state $l$, labelling this mask \lstinline|l_indices_contrib|.
For clarity we store the current $r$-realization's transition probability in \lstinline|temp|, to include in
the running product over chains.  If the current chain matches the chain of our derivative, we check that the
current $r$-realization (of $S_{t}^{\rm idx\_d}$) has state $k$ set; if so, we include in the product
for all realizations of $S_{t-1}^{\rm idx\_d}$, masked by those realizations having state $l$ set.  If we match the \lstinline|K-1| state, we need to take the derivative of the probability constraint on $A$.  Otherwise, this $A$ is not present so the derivative is zero.
In the final else statement, if we have not encountered the matching chain, we multiply-in the probability and continue looping.

The second derivative w.r.t $A$, written
$\partial^{2} \prod_{i} P(S_{t}^{i}|S_{t-1}^{i}) / \partial A_{mn}^{e} \partial A_{kl}^{d}$
is similar, but we need to check both sets of indices taking care if they are equal, and
apply the probability constraint, which contains more factors of $A$.
The code is as follows:
\begin{tcolorbox}
\begin{lstlisting}
d2prob_rdAdA *= temp[realizations[idx_d, :]]
d2prob_rdAdA = np.ones(shape=realizations.shape[1])
for idx_d in range(D):
    temp = np.exp(A)[idx_d, realizations[idx_d, r], :]
    if idx_d == d and d != e:
        if r in k_indices:
            d2prob_rdAdA *= temp[realizations[idx_d, :]]
        elif r in d_Km1_indices:
            temp2 = np.exp(A)[idx_d, k, :]
            d2prob_rdAdA *= -temp2[realizations[idx_d, :]]
        else:
            d2prob_rdAdA *= 0
            break

    elif idx_d == e and d != e:
        if r in m_indices:
            d2prob_rdAdA *= temp[realizations[idx_d, :]]
        elif r in e_Km1_indices:
            temp2 = np.exp(A)[idx_d, m, :]
            d2prob_rdAdA *= -temp2[realizations[idx_d, :]]
        else:
            d2prob_rdAdA *= 0  # The derivative is zero
            break

    elif idx_d == e and d == e:
        if r in k_indices and r in m_indices:
            d2prob_rdAdA *= temp[realizations[idx_d, :]] \ 
                            * l_indices_contrib \ 
                            * n_indices_contrib
        elif r in d_Km1_indices and r in e_Km1_indices:
            temp2 = np.exp(A)[idx_d, m, :]
            d2prob_rdAdA *= -temp2[realizations[idx_d, :]] \
                            * l_indices_contrib \
                            * n_indices_contrib
        else:
            d2prob_rdAdA *= 0
            break
    else:
        assert idx_d != d and idx_d != e
        d2prob_rdAdA *= temp[realizations[idx_d, :]]

\end{lstlisting}
\end{tcolorbox}
\noindent
In words similar to the single derivative: we first check that we have the same chain, masking the
derivative by $l$ realizations if this realization has $k$ set, if $K-1$ is set we apply the derivative of the constraint, otherwise zero; if the second derivative
has a different chain from the first, we mask the derivative by realizations contain $n$ if this
$r$-realization has $m$ set. If $d$ is $e$ we check that the term is in the product and the constraint, and apply the derivative. If neither of $d$ or $e$ match, we continue building the product as usual.
In the end, we should have non-zero entries that represent the appropriate
filtering of realizations by the two derivatives.

\subsubsection{$\pi$ derivatives}
The $\pi$ derivative only affects the initialization, which is a product of $\pi$'s, so the first
derivative omits that $\pi$ from the product or the constraint (with a minus sign), but if
$\pi$ is not in the product we get zero.  This is analogous to the $A$ calculation.  For example, the
code for the \lstinline|r|-realization value:
\begin{tcolorbox}
\begin{lstlisting}
# First derivative of joint pi
djoint_dpi = 1
for idx_d in range(D):
    if idx_d == e:
        if realizations[idx_d, i] == l:
            continue  # derivative implies excluding from product
        elif realizations[idx_d, i] == K-1:  # Last state by convention
            djoint_dpi *= -1
            continue
        else:
            djoint_dpi *= 0
    djoint_dpi *= pi[idx_d, realizations[idx_d, i]]

\end{lstlisting}
\end{tcolorbox}
For the second derivative, we need to check some index combinations, as with $A$.
\begin{tcolorbox}
\begin{lstlisting}
d2joint_dpidpi = 1
for idx_d in range(D):
    if idx_d == d and d != e:
        if realizations[idx_d, i] == k:
            mult_factor = 1
        elif realizations[idx_d, i] == K-1:  # Last state by convention
            mult_factor = -1
        else:
            d2joint_dpidpi *= 0
            break
    elif idx_d == e and d != e:
        if realizations[idx_d, i] == l:
            mult_factor = 1
        elif realizations[idx_d, i] == K-1:  # Last state by convention
            mult_factor = -1
        else:
            d2joint_dpidpi *= 0
            break
    elif idx_d == d and d == e:
        d2joint_dpidpi = 0  # These second derivatives will be zero
        break
    else:
        assert idx_d != d and idx_d != e
        mult_factor = pi[idx_d, realizations[idx_d, i]]

    d2joint_dpidpi *= mult_factor

\end{lstlisting}
\end{tcolorbox}

\subsubsection{Recursions}
Now that we have all the pieces we can look at the recursion calculations:

\begin{tcolorbox}
\begin{lstlisting}
alpha[t, r] = np.sum(alpha[t-1] * prob_r * py[t, r]) + eps
dalphadw[t, r] = \
    np.sum(alpha[t-1] * prob_r * dpydw[t, r]) \
    + np.sum((dalphadw[t-1] - dcdw[t-1]/c[t-1] * alpha[t-1]) * prob_r * py[t, r])
dalphadC[t, r] = \
    np.sum(alpha[t-1] * prob_r * dpydC[t, r]) \
    + np.sum((dalphadC[t-1] - dcdC[t-1]/c[t-1] * alpha[t-1]) * prob_r * py[t, r])
d2alphadwdC[t, r] = \
    np.sum((dalphadC[t-1] - alpha[t-1] * dcdC[t-1]/c[t-1]) * prob_r * dpydw[t, r]) \
    + np.sum(alpha[t-1] * prob_r * d2pydwdC[t, r]) \
    + np.sum((dalphadw[t-1] - dcdw[t-1]/c[t-1] * alpha[t-1]) * prob_r * dpydC[t, r]) \
    + np.sum((d2alphadwdC[t-1]
              + 2 * dcdC[t-1]/c[t-1] * dcdw[t-1]/c[t-1] * alpha[t-1]
              - d2cdwdC[t-1]/c[t-1] * alpha[t-1]
              - dcdw[t-1]/c[t-1] * dalphadC[t-1]
              - dcdC[t-1]/c[t-1] * dalphadw[t-1]) * prob_r * py[t, r])
\end{lstlisting}
\end{tcolorbox}
\noindent
These are the $\widetilde{\alpha}$ updates, where the sum in the assignment is the sum over $S_{t-1}$
configurations present in \lstinline|prob_r| (mathematically stemming from the term
$\sum_{\{S_{t-1}\}}\prod P(S_{t}|S_{t-1})$).
The first equality is the regular $\alpha$ update.  The next two are the first derivatives, where
we have discarded derivatives of \lstinline|prob_r|, since the product term doesn't depend on $W$ or $C$.
The final term is the second derivative which, when dropping the derivative of \lstinline|prob_r| terms,
only includes terms 1, 3, 7, and 9, not in that order.

We then sum over their realizations ($S_{t}$ configurations) to obtain and track $c$ and its derivatives,
as follows (for example):

\begin{tcolorbox}
\begin{lstlisting}
c[t] = alpha[t, :].sum()
alpha[t, :] /= c[t]                   # Normalize
dcdw[t] = dalphadw[t, :].sum()
dalphadw[t, :] /= c[t]                # Normalize
dcdC[t] = dalphadC[t, :].sum()
dalphadC[t, :] /= c[t]                # Normalize
d2cdwdC[t] = d2alphadwdC[t, :].sum()
d2alphadwdC[t, :] /= c[t]             # Normalize
\end{lstlisting}
\end{tcolorbox}
\noindent
We can then calculate the value of this Hessian element as follows:
\begin{tcolorbox}
\begin{lstlisting}
hessian_wc = 0
for t in range(T):
    hessian_wc += -1/c[t]**2 * dcdC[t] * dcdw[t] + 1/c[t] * d2cdwdC[t]
\end{lstlisting}
\end{tcolorbox}

To summarize, we have outlined all of the pieces that go into the Hessian element functions --
for example, 
\lstinline|hessian_WC(d, o, k, i, j)| which takes the three indices of $W$ and the two indices
of $C$ and returns the Hessian value.  With the corresponding functions for the other combinations
of $W$, $C$, $A$, $\pi$, we can loop over the all index combinations to create the full
Hessian matrix of shape $\mathrm{dim} \times \mathrm{dim}$ where
$\mathrm{dim} = dok - (d-1)o + d(k-1)k + o^2 + d(k-1)$.

}

%\bibliographystyle{unsrt}
%\bibliography{bibliography}

%
\begin{figure}[p!]
\begin{tabular}{c}
	\subfloat[$d$=2;$C$=0.0001]{\includegraphics[width=16cm, height=4cm]{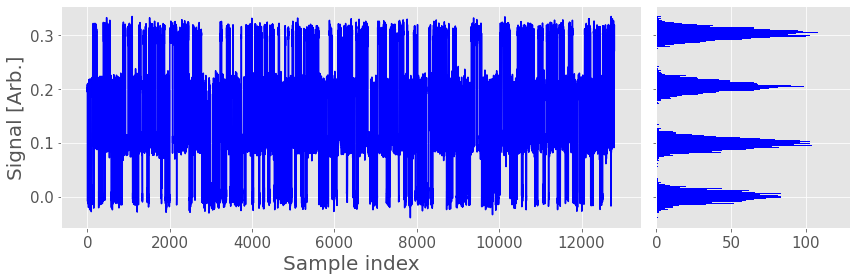}} \\
	\subfloat[$d$=2;$C$=0.01]{\includegraphics[width=16cm, height=4cm]{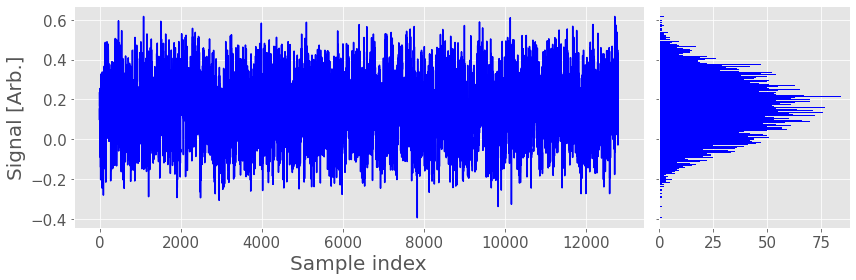}} \\
	\subfloat[$d$=4;$C$=0.0001]{\includegraphics[width=16cm, height=4cm]{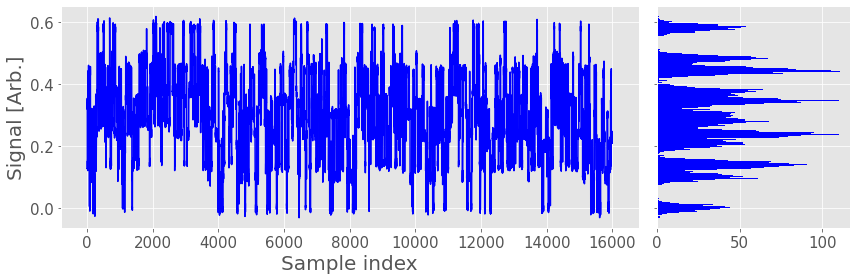}} \\
	\subfloat[$d$=4;$C$=0.01]{\includegraphics[width=16cm, height=4cm]{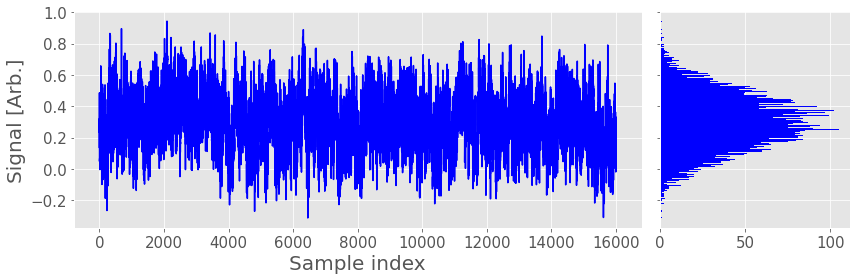}}
\end{tabular}
\caption{\textbf{FHMM time series examples.} These time series data, including histograms, showcase the visibility (or lack thereof) of discrete levels for each experiment.}
\label{fig:TShists}
\end{figure}
\begin{figure}[p!]
\begin{tabular}{cc}
	\subfloat[$d$=2;$C$=0.0001]{\includegraphics[width=8cm, height=6cm]{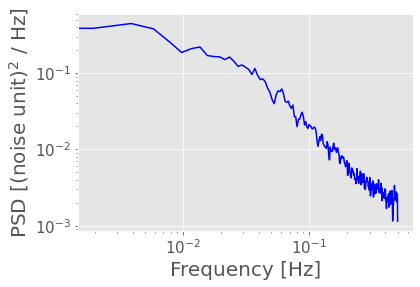}} &
	\subfloat[$d$=2;$C$=0.01]{\includegraphics[width=8cm, height=6cm]{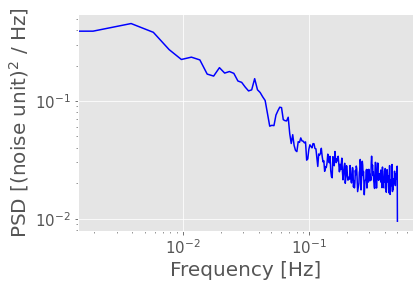}} \\
	\subfloat[$d$=4;$C$=0.0001]{\includegraphics[width=8cm, height=6cm]{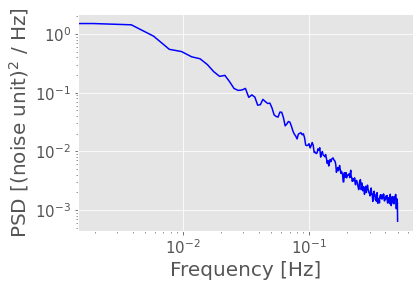}} &
	\subfloat[$d$=4;$C$=0.01]{\includegraphics[width=8cm, height=6cm]{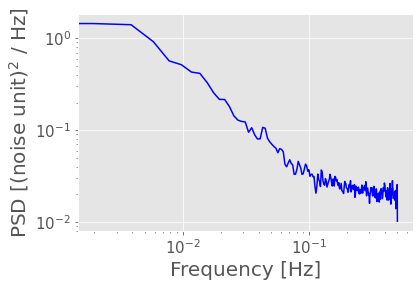}}
\end{tabular}
\caption{\textbf{FHMM PSD examples.} The power spectral densities for all experiments, highlighting the limited amount of information
present.}
\label{fig:PSDs}
\end{figure}
\begin{figure}[p!]
\begin{tabular}{ccc}
	\subfloat[]{\includegraphics[width=8cm, height=6cm]{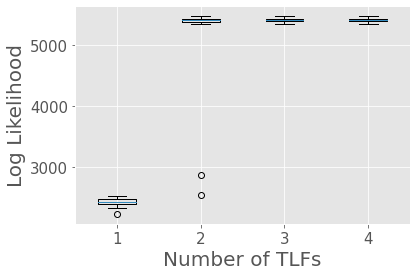}} &
	\subfloat[]{\includegraphics[width=8cm, height=6cm]{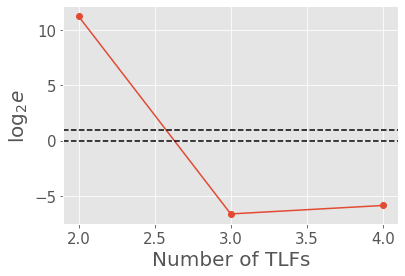}}
\end{tabular}
\caption{\textbf{FHMM CV results for $d$ = 2; $C$ = 0.0001.} (a) is the cross-validation results and
         (b) is the $\log_2$ of the evidence ratio,
         where each model corresponding to $d$ on the x-axis is compared with
         the $d-1$ model.  The region between the dashed lines is weak evidence for the
         larger model, while above the dashed lines is strong evidence, indicating
         $d=2$.}
\label{fig:cved2c0001}
\end{figure}
\begin{figure}[p!]
\begin{tabular}{ccc}
	\subfloat[]{\includegraphics[width=8cm, height=6cm]{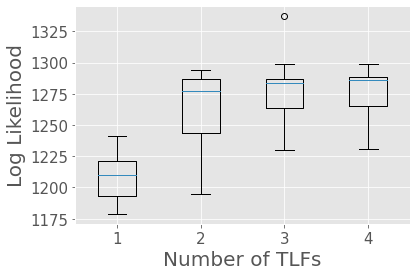}} &
	\subfloat[]{\includegraphics[width=8cm, height=6cm]{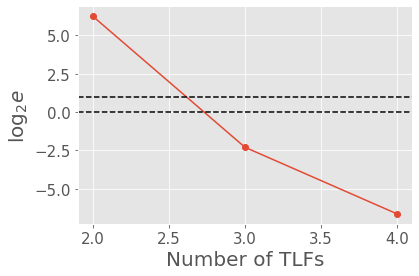}}
\end{tabular}
\caption{\textbf{FHMM CV results for $d$ = 2; $C$ = 0.01.} (a) is the cross-validation results and
         (b) is the $\log_2$ of the evidence ratio,
         where each model corresponding to $d$ on the x-axis is compared with
         the $d-1$ model.  The region between the dashed lines is weak evidence for the
         larger model, while above the dashed lines is strong evidence, indicating
         $d=2$.}
\label{fig:cved2c01}
\end{figure}
\begin{figure}[p!]
\begin{tabular}{ccc}
	\subfloat[]{\includegraphics[width=8cm, height=6cm]{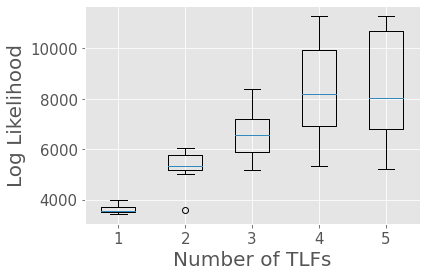}} &
	\subfloat[]{\includegraphics[width=8cm, height=6cm]{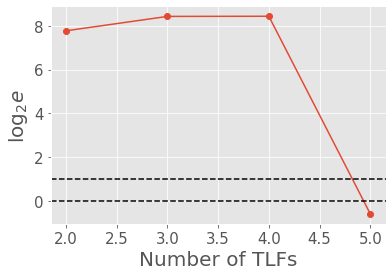}}
\end{tabular}
\caption{\textbf{FHMM CV results for $d$ = 4; $C$ = 0.0001.} (a) is the cross-validation results
         and (b) is the $\log_2$ of the evidence ratio,
         where each model corresponding to $d$ on the x-axis is compared with
         the $d-1$ model.  The region between the dashed lines is weak evidence for the
         larger model, while above the dashed lines is strong evidence, indicating
         $d=4$.}
\label{fig:cved4c0001}
\end{figure}
\begin{figure}[p!]
\begin{tabular}{ccc}
	\subfloat[]{\includegraphics[width=8cm, height=6cm]{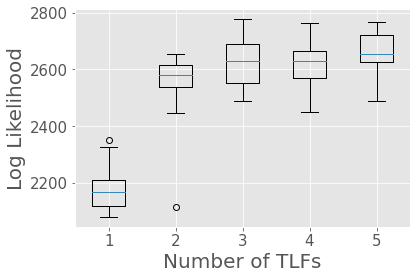}} &
	\subfloat[]{\includegraphics[width=8cm, height=6cm]{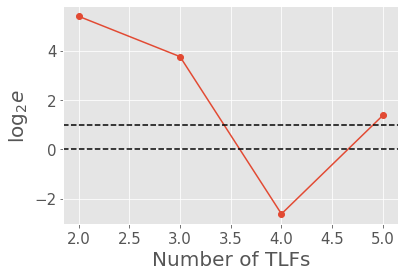}}
\end{tabular}
\caption{\textbf{FHMM CV results for $d$ = 4; $C$ = 0.01.} (a) is the cross-validation results
         and (b) is the $\log_2$ of the evidence ratio,
         where each model corresponding to $d$ on the x-axis is compared with
         the $d-1$ model.  The region between the dashed lines is weak evidence for the
         larger model, while above the dashed lines is strong evidence, indicating
         $d=3$.}
\label{fig:cved4c01}
\end{figure}
\begin{figure}[p!]
\begin{tabular}{ccc}
	\includegraphics[width=8cm, height=6cm]{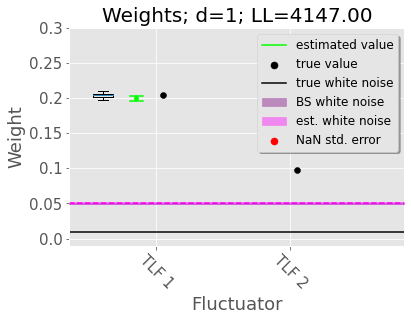} &
	\includegraphics[width=8cm, height=6cm]{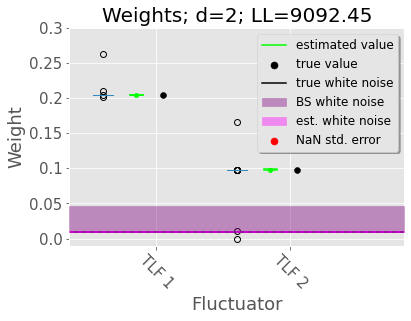} \\
	\includegraphics[width=8cm, height=6cm]{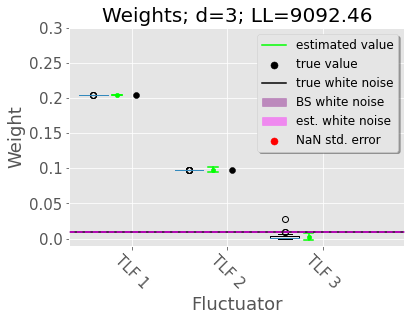} &
	\includegraphics[width=8cm, height=6cm]{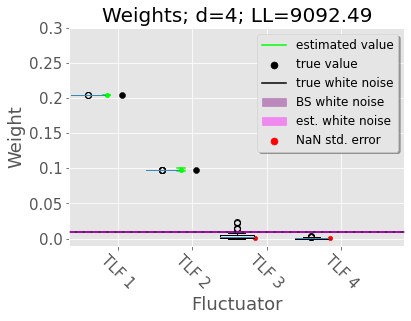}
\end{tabular}
\caption{\textbf{FHMM fitted weights for $d$=2 and $C$=0.0001.} The figures include
         weights for model fits with $d=1, 2, 3, 4$, Hessian-based confidence intervals
         at the $95\%$ confidence level (green error bars), as well as bootstrapped
         CI boxplots. The true, data-generating model has $d$=2 and $C$=0.0001.}
\label{fig:d2c0001W}
\end{figure}
\begin{figure}[p!]
\begin{tabular}{ccc}
	\includegraphics[width=8cm, height=6cm]{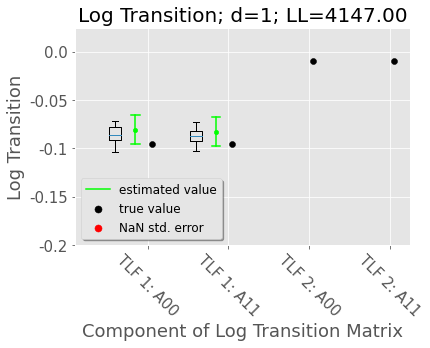} &
	\includegraphics[width=8cm, height=6cm]{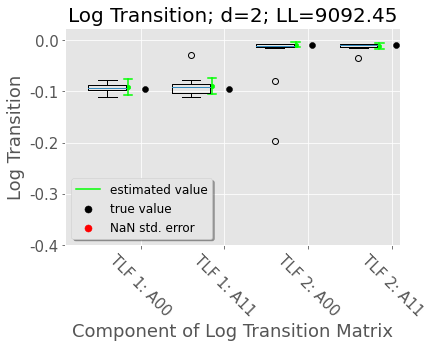} \\
	\includegraphics[width=8cm, height=6cm]{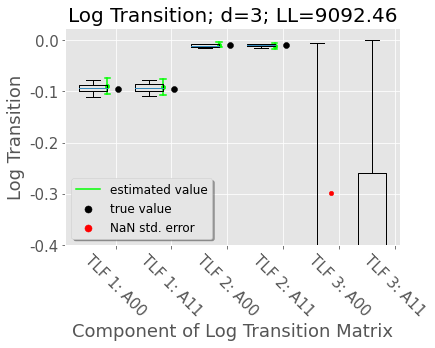} &
	\includegraphics[width=8cm, height=6cm]{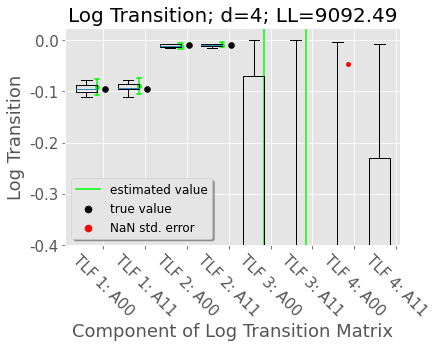}
\end{tabular}
\caption{\textbf{FHMM fitted log transitions for $d$=2 and $C$=0.0001.} The figures
         include log transitions for model fits with $d=1, 2, 3, 4$, including
         Hessian-based confidence intervals at the
	     $95\%$ confidence level (green error bars), as well as bootstrapped CI boxplots.
	     The true, data-generating model has $d$=2 and $C$=0.0001.}
\label{fig:d2c0001A}
\end{figure}
\begin{figure}[p!]
\begin{tabular}{ccc}
	\subfloat[$d$=1]{\includegraphics[width=8cm, height=6cm]{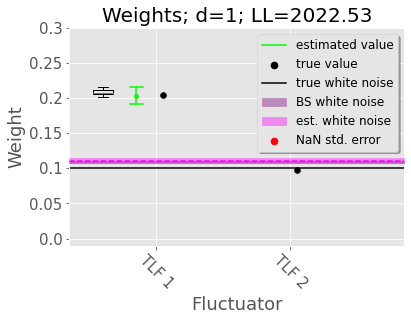}} &
	\subfloat[$d$=2]{\includegraphics[width=8cm, height=6cm]{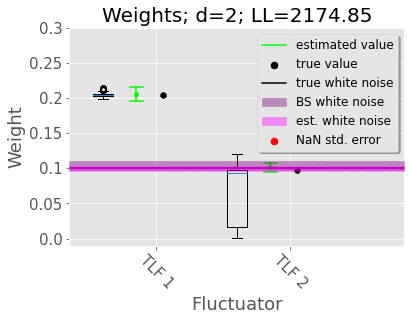}} \\
	\subfloat[$d$=3]{\includegraphics[width=8cm, height=6cm]{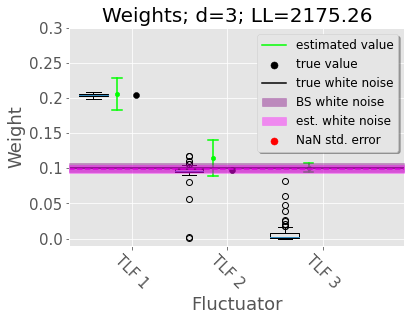}} &
	\subfloat[$d$=4]{\includegraphics[width=8cm, height=6cm]{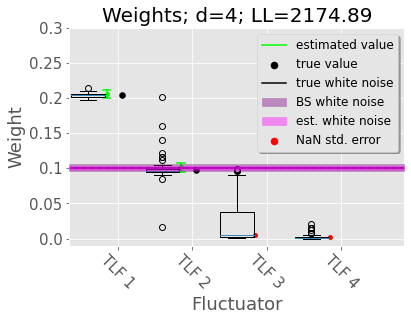}}
\end{tabular}
\caption{\textbf{FHMM fitted weights for $d$=2 and $C$=0.01.} The figures include weights for
         model fits with $d=1, 2, 3, 4$, Hessian-based confidence intervals at the
	     $95\%$ confidence level (green error bars), as well as bootstrapped CI boxplots.
	     The true, data-generating model has $d$=2 and $C$=0.01.}
\label{fig:d2c01W}
\end{figure}
\begin{figure}[p!]
\begin{tabular}{ccc}
	\subfloat[$d$=1]{\includegraphics[width=8cm, height=6cm]{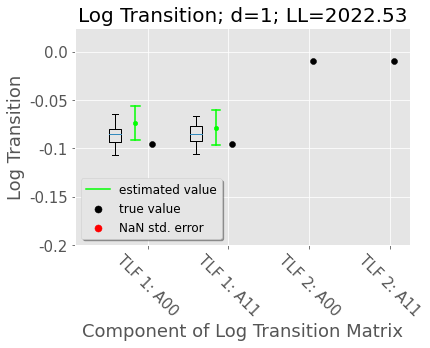}} &
	\subfloat[$d$=2]{\includegraphics[width=8cm, height=6cm]{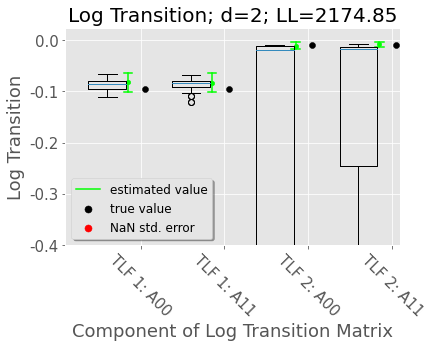}} \\
	\subfloat[$d$=3]{\includegraphics[width=8cm, height=6cm]{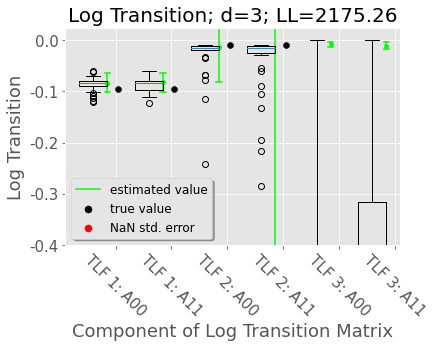}} &
	\subfloat[$d$=4]{\includegraphics[width=8cm, height=6cm]{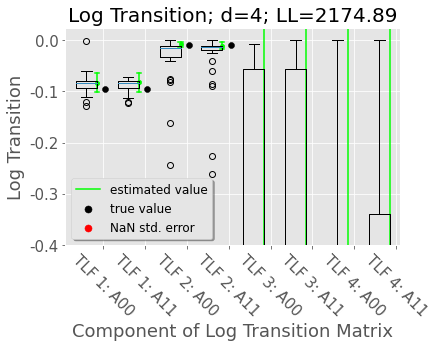}}
\end{tabular}
\caption{\textbf{FHMM fitted log transitions for $d$=2 and $C$=0.01.} The figures
         include log transitions for model fits with $d=1, 2, 3, 4$, including
         Hessian-based confidence intervals at the
	     $95\%$ confidence level (green error bars), as well as bootstrapped CI boxplots. The true,
	     data-generating model has $d$=2 and $C$=0.01.}
\label{fig:d2c01A}
\end{figure}
\begin{figure}[p!]
\begin{tabular}{ccc}
	\subfloat[$d$=1]{\includegraphics[width=8cm, height=6cm]{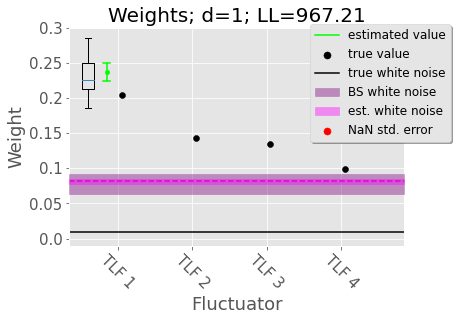}} &
	\subfloat[$d$=2]{\includegraphics[width=8cm, height=6cm]{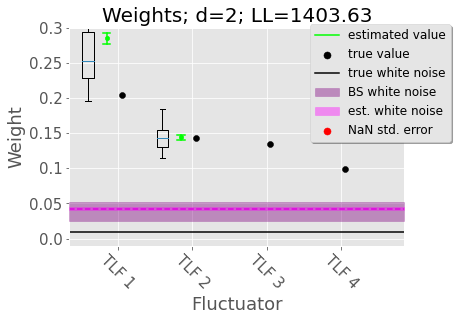}} \\
	\subfloat[$d$=3]{\includegraphics[width=8cm, height=6cm]{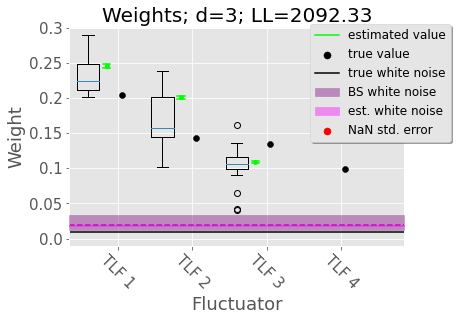}} &
	\subfloat[$d$=4]{\includegraphics[width=8cm, height=6cm]{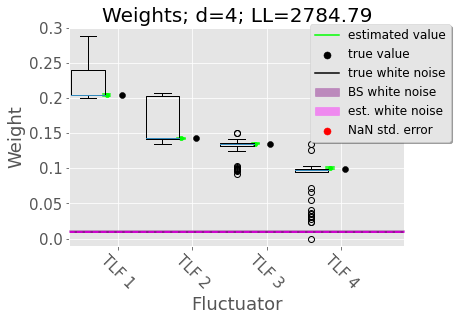}} \\
	\subfloat[$d$=5]{\includegraphics[width=8cm, height=6cm]{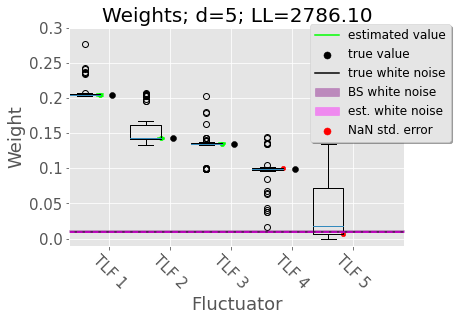}}
\end{tabular}
\caption{\textbf{FHMM fitted weights for $d$=4 and $C$=0.0001.} The figures
         include weights for model fits with $d=1, 2, 3, 4, 5$, Hessian-based
         confidence intervals at the
	     $95\%$ confidence level (green error bars), as well as bootstrapped CI boxplots. The true, data-generating model has $d$=4 and $C$=0.0001.}
\label{fig:d4c0001W}
\end{figure}
\begin{figure}[p!]
\begin{tabular}{ccc}
	\subfloat[$d$=1]{\includegraphics[width=8cm, height=6cm]{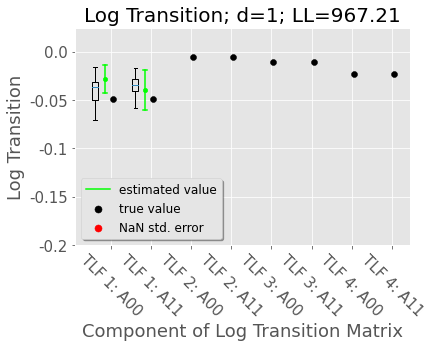}} &
	\subfloat[$d$=2]{\includegraphics[width=8cm, height=6cm]{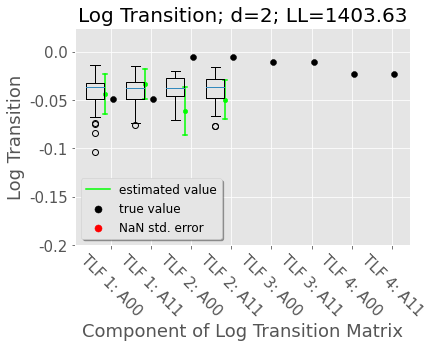}} \\
	\subfloat[$d$=3]{\includegraphics[width=8cm, height=6cm]{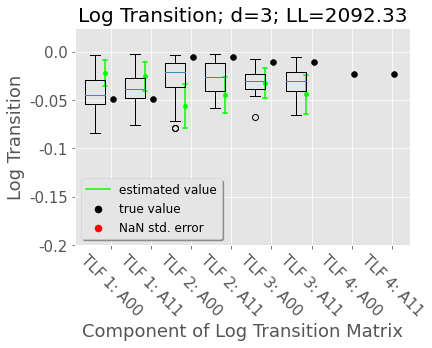}} &
	\subfloat[$d$=4]{\includegraphics[width=8cm, height=6cm]{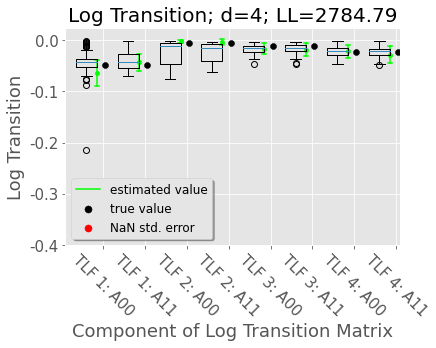}} \\
	\subfloat[$d$=5]{\includegraphics[width=8cm, height=6cm]{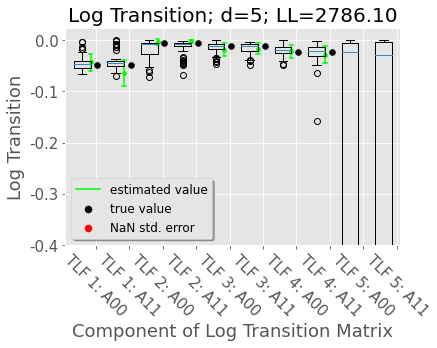}}
\end{tabular}
\caption{\textbf{FHMM fitted log transitions for $d$=4 and $C$=0.0001.} The figures
         include log transitions for model fits with $d=1, 2, 3, 4, 5$, including
         Hessian-based confidence intervals at the
	     $95\%$ confidence level (green error bars), as well as bootstrapped CI boxplots. The true,
	     data-generating model has $d$=4 and $C$=0.0001.}
\label{fig:d4c0001A}
\end{figure}
\begin{figure}[p!]
\begin{tabular}{ccc}
	\subfloat[$d$=1]{\includegraphics[width=8cm, height=6cm]{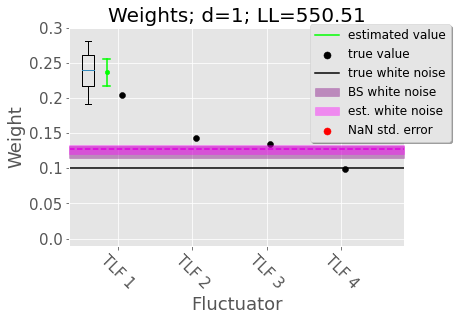}} &
	\subfloat[$d$=2]{\includegraphics[width=8cm, height=6cm]{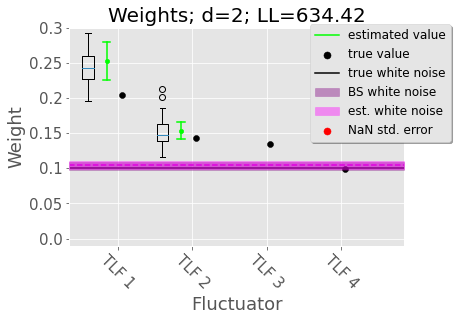}} \\
	\subfloat[$d$=3]{\includegraphics[width=8cm, height=6cm]{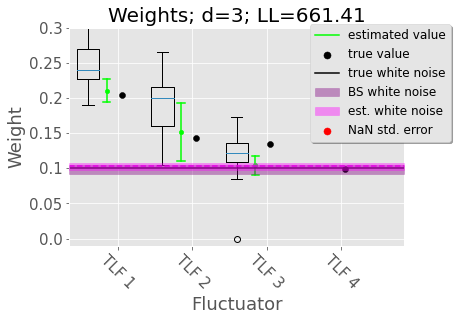}} &
	\subfloat[$d$=4]{\includegraphics[width=8cm, height=6cm]{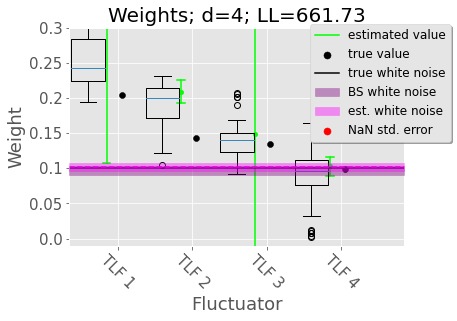}} \\
	\subfloat[$d$=5]{\includegraphics[width=8cm, height=6cm]{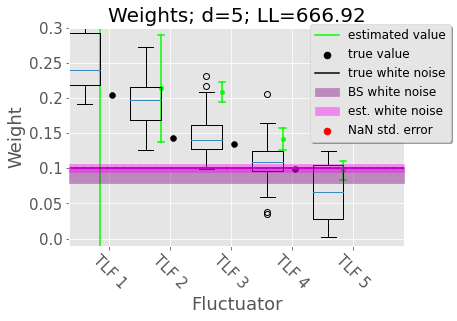}}
\end{tabular}
\caption{\textbf{FHMM fitted weights for $d$=4 and $C$=0.01.} The figures
         include weights for model fits with $d=1, 2, 3, 4, 5$, Hessian-based
         confidence intervals at the
	     $95\%$ confidence level (green error bars), as well as bootstrapped CI boxplots. The true, data-generating model has $d$=4 and $C$=0.01.}
\label{fig:d4c01W}
\end{figure}
\begin{figure}[p!]
\begin{tabular}{ccc}
	\subfloat[$d$=1]{\includegraphics[width=8cm, height=6cm]{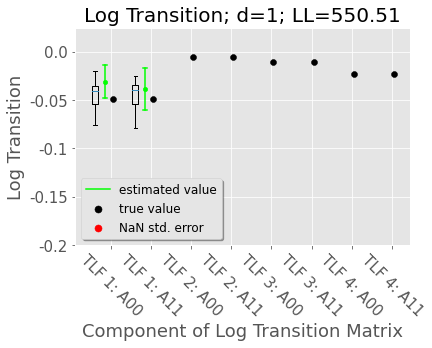}} &
	\subfloat[$d$=2]{\includegraphics[width=8cm, height=6cm]{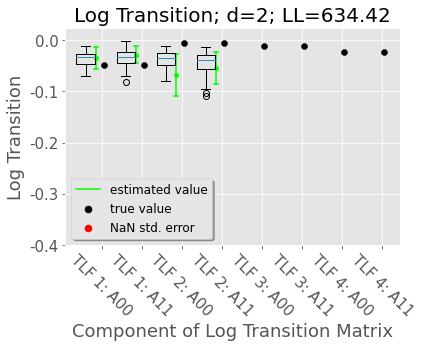}} \\
	\subfloat[$d$=3]{\includegraphics[width=8cm, height=6cm]{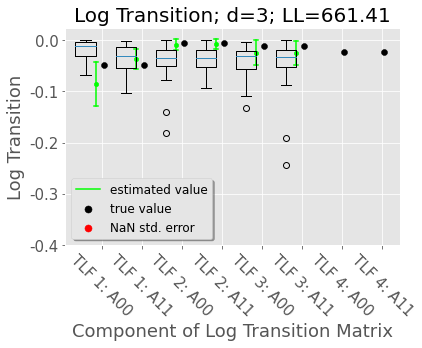}} &
	\subfloat[$d$=4]{\includegraphics[width=8cm, height=6cm]{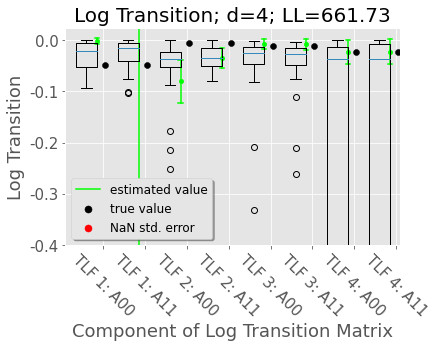}} \\
	\subfloat[$d$=5]{\includegraphics[width=8cm, height=6cm]{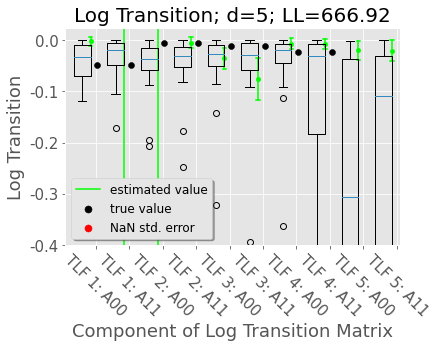}}
\end{tabular}
\caption{\textbf{FHMM fitted log transitions for $d$=4 and $C$=0.01.} The figures
         include log transitions for model fits with $d=1, 2, 3, 4, 5$, including
         Hessian-based confidence intervals at the
	     $95\%$ confidence level (green error bars), as well as bootstrapped CI boxplots. The true,
	     data-generating model has $d$=4 and $C$=0.01.}
\label{fig:d4c01A}
\end{figure}
\begin{figure}[p!]
\begin{tabular}{c}
	\subfloat[$d$=4 TLF system with noise.  First 100k samples out of 10M points.]{\includegraphics[width=16cm, height=6cm]{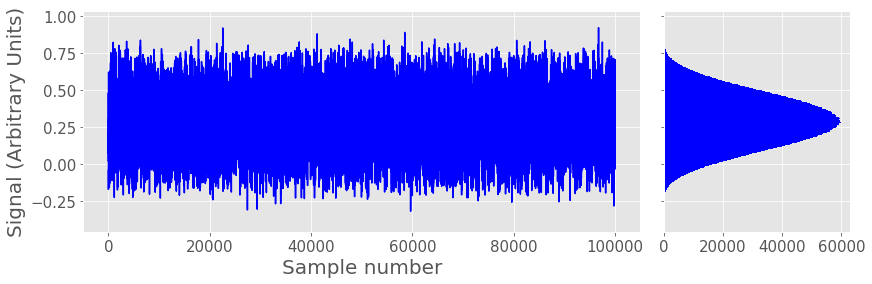}} \\
	\subfloat[Gaussian $1/f^{\beta}$ noise.  First 100k samples out of 20M points.]{\includegraphics[width=16cm, height=6cm]{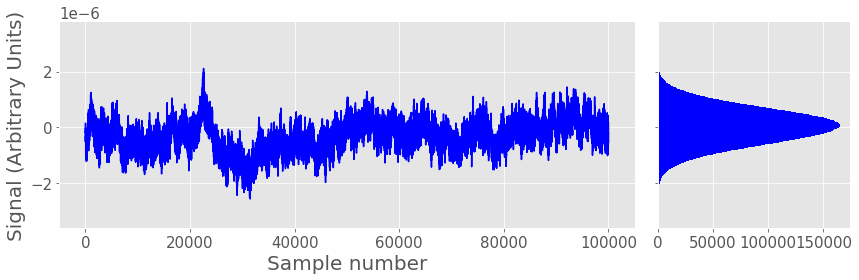}}
\end{tabular}
\caption{\textbf{Higher order statistics time series examples.} (a) TLF and
         (b) $1/f^{\beta}$ time series traces and histograms for the two examples used
         in the second spectrum analysis.}
\label{fig:hosTS}
\end{figure}

\begin{figure}[p!]
\begin{tabular}{cc}
	\subfloat[$d$=4 TLF system PSD.]{\includegraphics[width=8cm, height=6cm]{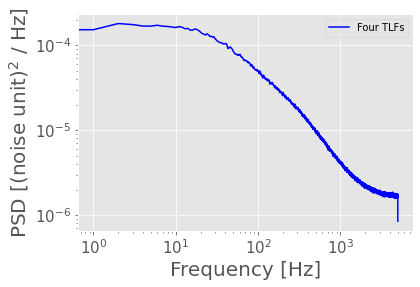}} &
	\subfloat[Gaussian $1/f^{\beta}$ noise PSD.]{\includegraphics[width=8cm, height=6cm]{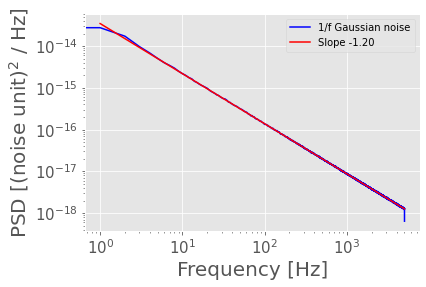}}
\end{tabular}
\caption{\textbf{Higher order statistics PSD examples.} (a) TLF and (b) $1/f^{\beta}$
         power spectral densities of example data used in the second spectrum analysis.
         Each displays distinct $1/f$ behavior.}
\label{fig:hosPSD}
\end{figure}

\begin{figure}[p!]
\begin{tabular}{cc}
	\subfloat[$d$=4 TLF system second spectrum.]{\includegraphics[width=8cm, height=6cm]{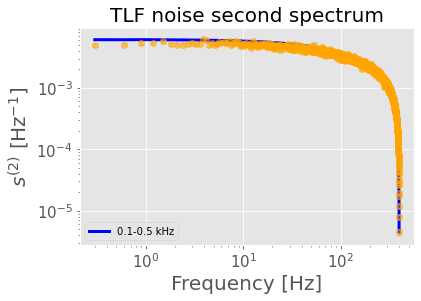}} &
	\subfloat[Gaussian $1/f^{\beta}$ noise second spectrum.]{\includegraphics[width=8cm, height=6cm]{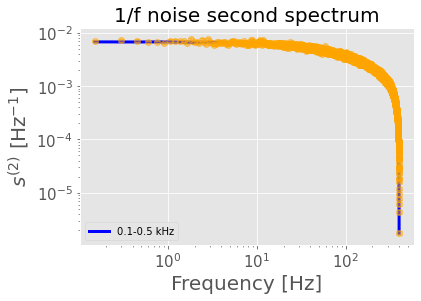}}
\end{tabular}
\caption{\textbf{Higher order statistics $s^{(2)}$ examples.} (a) TLF and
         (b) $1/f^{\beta}$ second spectra for the 0.1-0.5 kHz band.  The solid line on
         each figure is the Gaussian background second spectrum. }
\label{fig:hosSecondSpectrum}
\end{figure}

\pagebreak

% a la SymPy
\flushbottom
\thispagestyle{empty}%
\vskip-36pt%
{\raggedright\sffamily\bfseries\fontsize{20}{25}\selectfont NoMoPy: Noise Modeling in Python \par}%
\vskip10pt
{\raggedright\sffamily\fontsize{12}{16}\selectfont  Supplementary material\par}
\vskip25pt%

Here we include the derivations and implementations of many of the algorithms
present in \nomopy.

\bibliographystyle{unsrt}
\bibliography{bibliography}

\end{document}